\documentstyle[prd,aps,epsfig,preprint,tighten]{revtex}
\begin{document}
\draft
\title{		Quasi-vacuum solar neutrino oscillations}
\author{	G.L.\ Fogli$^a$, 
		E.\ Lisi$^a$, 
		D.\ Montanino$^b$, and 
		A.\ Palazzo$^a$}
\address{     	$^a$~Dipartimento di Fisica and Sezione INFN di Bari,\\
               	Via Amendola 173, I-70126 Bari, Italy}
\address{	$^b$~Dipartimento di Scienza dei Materiali 
		dell'Universit\`a di Lecce,\\
             	Via Arnesano, I-73100 Lecce, Italy}
\maketitle
\begin{abstract}
We discuss in detail solar neutrino oscillations with $\delta m^2/E$ in the range
$[10^{-10},10^{-7}]$ eV$^2$/MeV. In this range, which interpolates  smoothly between
the so-called ``just-so''  and ``Mikheyev-Smirnov-Wolfenstein'' oscillation regimes,
neutrino flavor transitions are increasingly affected by matter effects as $\delta
m^2/E$ increases.  As a consequence, the usual vacuum approximation has to be
improved through the  matter-induced corrections, leading to a ``quasi-vacuum''
oscillation regime. We perform accurate numerical calculations of such corrections,
using both the true solar density profile and its exponential approximation. Matter
effects  are shown to be somewhat overestimated in the latter case. We also discuss
the role of Earth crossing and of energy smearing. Prescriptions are given to 
implement the leading corrections in the quasi-vacuum oscillation range. Finally, the
results are applied to a global analysis of solar $\nu$  data in a three-flavor
framework.
\end{abstract}
\pacs{\\ PACS number(s): 26.65.+t, 14.60.Pq}

\section{Introduction}
\label{S1}

A well-known explanation of the solar $\nu_e$ flux deficit \cite{NuAs} is provided by
flavor oscillations \cite{Po67} of neutrinos along their way from the Sun $(\odot)$
to the Earth $(\oplus)$. For two active neutrino states [say, $(\nu_e,\nu_\mu)$ in
the flavor basis and $(\nu_1,\nu_2)$ in the mass basis], the physics of solar $\nu$
oscillations is governed, at any given energy $E$, by  the mass-mixing parameters
$\delta m^2$ and $\omega$ in vacuum,%
\footnote{One can take $\delta m^2=m^2_2-m^2_1>0$, as far as $\omega$ is taken
in the first quadrant $[0,\frac{\pi}{2}]$.}
as well as by the electron density profile $N_e(x)$ in matter \cite{MSWm}.

Different oscillation regimes can be identified in terms of three
characteristics lengths, namely, the astronomical unit
\begin{equation}
L=1.496\times 10^8\ {\rm\ km}\ ,
\label{L}
\end{equation}
the oscillation length in vacuum
\begin{equation}
L_{\rm osc}=\frac{4\pi E}{\delta m^2} = 2.48\times 10^{-3}\left( 
\frac{\delta m^2/E}{{\rm eV}^2/{\rm MeV}} \right)^{-1}\;{\rm km}\ ,
\label{Losc}
\end{equation}
and the refraction length in matter
\begin{equation}
L_{\rm mat} = \frac{2\pi}{\sqrt{2}G_F N_e}= 1.62\times 10^4
\left( \frac{N_e}{{\rm mol}/{\rm cm}^3} \right)^{-1}\;{\rm km}\ ,
\label{Lmat}
\end{equation}
which is associated to the effective mixing angle $\omega_m$ \cite{MSWm},
\begin{equation}
\sin^2 2\omega_m = \frac{\sin^2 2\omega}{(\cos 2\omega - 
L_{\rm osc}/L_{\rm mat})^2
+\sin^2 2\omega}\ .
\label{omegam}
\end{equation}
Typical  solutions to the solar neutrino problem (see, e.g., \cite{300d})
involve  values of $L_{\rm osc}$ either in the so-called ``just-so'' (JS)
oscillation regime \cite{Gl87}, characterized by
\begin{equation}
 L^{\rm JS}_{\rm osc} \sim L \gg L_{\rm mat}\ ,
\label{JS}
\end{equation}
 or in the  ``Mykheyev-Smirnov-Wolfenstein'' (MSW) oscillation regime
\cite{MSWm},  characterized by 
\begin{equation}
L_{\rm osc}^{\rm MSW} \sim L_{\rm mat}  \ll L\ .
\label{MSW}
\end{equation}
The two regimes correspond roughly to $\delta m^2/E\sim O(10^{-11})$ eV$^2$/MeV
and to $\delta m^2/E \gtrsim 10^{-7}$ eV$^2$/MeV, respectively.

For just-so oscillations, since $L_{\rm mat}/L_{\rm osc}\to 0$, the effect of matter
is basically to suppress the oscillation amplitude both in the Sun and in the Earth 
($\sin^2 2\omega_m\to 0$), so that (coherent) flavor oscillations take place only in
vacuum, starting from the Sun surface \cite{Kr85}. Conversely, for MSW oscillations,
$L_{\rm osc}\sim L_{\rm mat}$ and flavor transitions are dominated by the detailed
matter density profile, while the many oscillation cycles in vacuum 
$(L_{\rm osc}\ll L)$ are responsible for complete $\nu$ decoherence at the Earth, once
smearing effects  are taken into account \cite{Cohe}.

Therefore, it is intuitively
clear that in the intermediate range between (\ref{JS}) and (\ref{MSW}),
corresponding approximately to $10^{-10}\lesssim \delta m^2/E\lesssim 10^{-7}$
eV$^2$/MeV, the simple vacuum oscillation picture of the JS regime becomes
increasingly decoherent and affected by matter effects for increasing values of
$\delta m^2/E$, leading to a hybrid regime that might be called of
``quasi-vacuum'' (QV) oscillations, characterized by
\begin{equation}
L_{\rm mat}\lesssim L^{\rm QV}_{\rm osc}\lesssim L\ ,
\end{equation}

In the past, quasi-vacuum effects on the oscillation amplitude and phase have been
explicitly considered only in relatively few papers (see, e.g.,
\cite{To87,Ka87,Pe88,PeRi,Pa90,Pa91,Ba96}) as compared with the vast literature on
solar neutrino oscillations,  essentially because typical fits to solar $\nu$ rates
allowed  only marginal solutions in the range where QV effects are relevant.
However, more recent analyses appear to extend the former ranges of the JS solutions
{\em upwards\/}  \cite{Ba00} and of the MSW solutions {\em downwards\/} \cite{Fo99}
in $\delta m^2/E$, making them eventually merge in the QV range \cite{Go00},
especially under generous assumptions about the  experimental or theoretical $\nu$
flux uncertainties.%
\footnote{Such extended range for current 
solutions reflects, in part, the lack of a strong, model-independent
signature of energy dependence in the solar neutrino deficit.}
Therefore, a fresh look at QV corrections seems warranted. Recently, a
semianalytical approximation improving the familiar just-so formula in the QV
regime was discussed in \cite{Fr99} and, in more detail, in \cite{Fr00}, where
additional numerical checks were performed. In this work we revisit the whole
topic, by performing accurate numerical calculations which include the exact 
density profile in the Sun and in the Earth, within the reference  mass-mixing
ranges $\delta m^2/E\in [10^{-10},10^{-7}]$ eV$^2$/MeV and $\tan^2\omega\in
[10^{-3},10]$.%
\footnote{ The variable $\tan^2\omega$ is useful to chart the first two octants
of the mixing angle range \protect\cite{Fo99,Fo96}.} 
We also discuss some approximations that can simplify the computing task in
present applications. We then apply such calculations to a global analysis of
solar neutrino data in the range $\delta m^2\leq 10^{-8}$ eV$^2$.

Our paper is structured as follows. The basic notation and the numerical
techniques used in the calculations are introduced in  Sec.~\ref{S2} and
\ref{S3}, respectively. The effects of solar matter in the quasi-vacuum
oscillation regime are discussed in Sec.~\ref{S4}, where the results for true
and exponential density profiles are compared. Earth matter effects are
described in Sec.~\ref{S5}. The decoherence of oscillations induced by energy
(and time) integration is discussed in Sec.~\ref{S6}.  The basic results are
summarized and organized in Sec.~\ref{S7}, and then applied to a three-flavor
oscillation analysis in Sec.~\ref{S8}. Section~\ref{S9} concludes our work.

\section{Notation}
\label{S2}

The $\nu$ propagation from the Sun core to the detector at the Earth can be
interpreted as a ``double slit experiment,'' where the original $\nu_e$ can
take two paths, corresponding to the intermediate  transitions $\nu_e\to\nu_1$
and to $\nu_e\to\nu_2$. The global $\nu_e$ survival amplitude is then the sum
of the amplitudes along the two paths,
\begin{eqnarray}
A(\nu_e\to\nu_e) 
&=& A_\odot(\nu_e\to\nu_1)\cdot A_{\rm vac}(\nu_1\to\nu_1)\cdot 
A_\oplus(\nu_1\to\nu_e) \nonumber\\
&+&  A_\odot(\nu_e\to\nu_2)\cdot A_{\rm vac}(\nu_2\to\nu_2)\cdot 
A_\oplus(\nu_2\to\nu_e)\ ,
\label{2path}
\end{eqnarray}
where the transition amplitudes from the Sun production point to its surface
$(A_\odot)$, from the Sun surface to the Earth surface $(A_{\rm vac})$ and from
the Earth surface to the detector $(A_\oplus)$ have been explicitly factorized.
The $\nu_e$ survival probability $P_{ee}$ is then given by
\begin{equation}
P_{ee}= |A(\nu_e\to\nu_e)|^2\ .
\label{Psurv}
\end{equation}

In general, the above amplitudes can be written as
\begin{mathletters}
\begin{eqnarray}
A_\odot(\nu_e\to\nu_1) 		&=& \sqrt{P_\odot}\exp(i\xi_\odot)\ ,\label{Av11}\\
A_{\rm vac}(\nu_1\to\nu_1) 	&=& 
\exp\left(-i\,m^2_1\,(L-R_\odot)/2E\right)\ ,\label{Av12}\\
A_\oplus(\nu_1\to\nu_e) 	&=& \sqrt{P_\oplus}\exp(i\xi_\oplus)\ ,\label{Av13}
\end{eqnarray}
\end{mathletters}
for the first path and as
\begin{mathletters}
\begin{eqnarray}
A_\odot(\nu_e\to\nu_2) 		&=& \sqrt{1-P_\odot}\ ,\label{Av21}\\
A_{\rm vac}(\nu_2\to\nu_2) 	&=& 
\exp\left(-i\,m^2_2\,(L-R_\odot)/2E\right)\ ,\label{Av22}\\
A_\oplus(\nu_2\to\nu_e) 	&=& \sqrt{1-P_\oplus}\ ,\label{Av23}
\end{eqnarray}
\end{mathletters}
for the second path, where $R_\odot$ is the Sun radius.%
\footnote{Although $R_\odot$ is relatively small  ($R_\odot/L=4.7\times
10^{-3}$), it is explicitly kept for later purposes. The Earth radius
$R_\oplus$ can instead be safely neglected ($R_\oplus/L=4.3\times 10^{-5}$).}
In the above equations, $P_\odot$ and $P_\oplus$ are real numbers ($\in[0,1]$)
representing the transition probability $P(\nu_e\leftrightarrow\nu_1)$ along
the two  partial paths inside the Sun (up to its surface) and inside the Earth
(up to the detector). The corresponding phase differences between the two
paths, $\xi_\odot$ and $\xi_\oplus$ $(\in [0,2\pi])$, have been associated to
the first path without loss of generality. The $\nu_e$ survival probability
$P_{ee}$ from Eq.~(\ref{Psurv}) reads then
\begin{eqnarray}
P_{ee} &=&
P_\odot P_\oplus + (1-P_\odot)(1-P_\oplus)\nonumber\\
& &+2\sqrt{P_\odot(1-P_\odot)P_\oplus(1-P_\oplus)}\cos\xi\ ,
\label{Pmat}
\end{eqnarray}
where the global oscillation phase is given by
\begin{equation}
\xi = \frac{\delta m^2
L}{2E}(1-\delta_R-\delta_\odot-\delta_\oplus)\ ,
\end{equation}
with the definitions
\begin{eqnarray}
\delta_{R} &=& \frac{R_\odot}{L}=4.7\times 10^{-3}\ ,\\
\delta_{\odot} &=& \frac{2E}{\delta m^2\, L}\xi_{\odot}\ ,\\
\delta_{\oplus} &=& \frac{2E}{\delta m^2\, L}\xi_{\oplus}\ .
\end{eqnarray}

A relevant extension of the $2\nu$ formula (\ref{Pmat}) is obtained for $3\nu$
oscillations, required to accommodate solar and atmospheric neutrino data
\cite{3atm}. Assuming a third mass eigenstate $\nu_3$ with
$m^2=|m^2_3-m^2_{1,2}|\gg\delta m^2$, the $3\nu$ survival probability can be
written as
\begin{equation}
P_{ee}^{3\nu} = c^4_\phi P_{ee}^{2\nu} + s^4_\phi\ , 
\label{P3nu}
\end{equation}
where $\phi$ is the $(\nu_e,\nu_3)$ mixing angle, and $P^{2\nu}_{ee}$ is given
by Eq.~(\ref{Pmat}), {\em provided that\/} the electron density $N_e$ is
replaced everywhere by $c^2_\phi\,N_e$ (see \cite{Fo96}  and refs.\ therein). 
Such replacement implies that the $3\nu$ case is not a simple mapping of the
$2\nu$ case, and requires specific calculations for any given value of $\phi$.

We conclude this section by recovering some familiar expressions for $P_{ee}$,
as special cases of Eq.~(\ref{Pmat}).  The JS limit ($L_{\rm mat}/L_{\rm
osc}\to 0$, with complete suppression of oscillations inside matter)
corresponds to $P_\odot \simeq c^2_\omega \simeq P_\oplus$ and to negligible
$\delta_\odot$, $\delta_\oplus$. Then, neglecting also 
$\delta_R$, one gets from
(\ref{Pmat}) the standard ``vacuum oscillation formula,''
\begin{equation}
P_{ee}^{\rm JS} \simeq 1-\sin^22\omega\sin^2(\pi L/L_{\rm osc})\ .
\label{Pvac}
\end{equation}
In the MSW limit ($L/L_{\rm osc}\to\infty$), the global oscillation phase $\xi$
is very large and $\cos\xi\simeq 0$ on average. Furthermore, assuming for
$P_\odot$ a well-known approximation in terms of the ``crossing'' probability
$P_c$ between mass eigenstates in matter  [in our notation,
$P_\odot\simeq\sin^2 \omega_m^0 P_c+\cos^2\omega_m^0(1-P_c)$, with $\omega_m^0$
calculated at the production point], one gets from (\ref{Pmat}) 
and for daytime $(P_\oplus= c^2_\omega)$ the so-called
Parke's formula \cite{Pa86}\ ,
\begin{equation}
P_{ee,\,{\rm day}}^{\rm MSW} \simeq \textstyle\frac{1}{2} + (\textstyle\frac{1}{2} - P_c)
\cos2\omega \cos 2\omega_m^0\ .
\end{equation}

\section{Numerical Techniques}
\label{S3}

In general,  numerical calculations of the $\nu$ transition amplitudes must 
take into account the detailed $N_e$ profile along the neutrino trajectory,
both in the Sun and in the Earth.

Concerning the Sun, we take $N_e$ from \cite{BaPi} (``year 2000'' standard
solar model). Figure~\ref{f1} shows such $N_e$ profile  as a function of the
normalized radius $r/R_\odot$, together with its exponential approximation
\cite{NuAs}  $N_e = N_e^0 \exp (-r/r_0)$,   with $N_e^0 =245$~mol/cm$^3$ and
$r_0=R_\odot/10.54$. For the exponential density profile, the neutrino
evolution equations can be solved analytically \cite{To87,Ka87,Pe88,PeRi}. In
order to calculate the relevant probability $P_\odot$ and the phase
$\xi_\odot$, we have developed several computer programs which evolve
numerically the familiar MSW neutrino evolution equations \cite{MSWm}  along
the Sun radius, for generic production points, and for any given value of
$\delta m^2/E\in [10^{-10},10^{-7}]$ eV$^2$/MeV and of $\tan^2\omega$. We
estimate  a numerical (fractional) accuracy of our results better than
$10^{-4}$, as derived by several independent checks. As a first test, we
integrate numerically the MSW equations both in their usual complex form (2
real + 2 imaginary components) and in their Bloch form involving three real
amplitudes \cite{Bloc}, obtaining the same results. We have then repeated the
calculations with different integration routines taken from several computer
libraries, and found no significant differences among the outputs.  We have
optionally considered, besides the exact $N_e$ profile, also the
exponential profile, which allows a further comparison of the numerical
integration of the MSW equations with their analytical solutions, as worked out
in  \cite{Pe88,PeRi} in terms of hypergeometric functions  (that we have
implemented in an independent code).  Also in this case, no difference is found
between the output of the different codes.

Concerning the calculation of the quantities $P_\oplus$ and $\xi_\oplus$ in
the Earth, we evolve analytically the MSW equations at any given nadir angle
$\eta$, using the technique described in \cite{Li97}, which is based on a 
five-step biquadratic approximation of the density profile from the 
Preliminary Reference Earth Model (PREM)  \cite{PREM}  and on a first-order
perturbative expansion of the neutrino evolution operator. Such analytical
technique provides results very close to a full numerical evolution of the
neutrino amplitudes,  the differences  being smaller than those induced by
uncertainties in $N_e$ \cite{Li97}. In particular, we have checked that, for
$\delta m^2/E\leq 10^{-7}$ eV$^2$/MeV, such differences are $\lesssim
10^{-3}$.  In conclusion, we are confident in the accuracy of our results,
which are discussed in the following sections.

\section{Matter effects in the Sun}
\label{S4}

Figure~\ref{f2} shows, in the mass-mixing plane and for standard solar model
density, isolines of the difference $c^2_\omega-P_\odot$ (solid curves), which
becomes zero in the just-so oscillation limit of very small $\delta m^2/E$. The
isolines shape reminds the ``lower corner'' of the more familiar MSW triangle
\cite{Pa86}. Also shown are isolines of constant resonance radius $R_{\rm
res}/R_{\odot}$ (dotted curves),  defined by the  MSW resonance condition
$L_{\rm osc}/L_{\rm mat}(R_{\rm res})= \cos {2\omega}$. The values of
$c^2_\omega-P_\odot$ are already sizable (a few percent) at $\delta m^2/E\sim
10^{-9}$, and increase for increasing $\delta m^2/E$ and for large mixing
$[\tan^2\omega\sim O(1)]$, especially in the first octant, where the MSW
resonance can occur. The difference between matter effects in the first and in
the second octant  can lead to observable modifications of the allowed regions
in fits to the data \cite{Fr00}, and  to a possible discrimination between the
cases $\omega<\frac{\pi}{4}$ and $\omega>\frac{\pi}{4}$  \cite{Go00,Fr00}.

In the whole parameter range of Fig.~\ref{f2}, it turns out that, within the
region of $\nu$ production ($r/R_\odot\lesssim 0.3$), it is  $L_{\rm mat}(r)\ll
L_{\rm osc}$ (and thus $\sin^2 2\omega_m^0\simeq 0$).%
\footnote{This is also indicated by the fact that 
$R_{\rm res}/R_\odot \gtrsim 0.55$ in the $\delta m^2/E$ range of Fig.~\ref{f2}.}
As a consequence, all the curves of Fig.~\ref{f2} do not depend on the specific
$\nu$ production point   (as we have also checked numerically), and no smearing
over the $\nu$ source distribution is needed in the quasi-vacuum regime.  This
is a considerable  simplification with respect to the MSW regime, which
involves higher values of $\delta m^2/E$  and thus shorter (resonance) radii,
which are sensitive to  the detailed $\nu$ source  distribution.

Figure~\ref{f3} shows, in the same coordinates of Fig.~\ref{f2}, the isolines
of $c^2_\omega-P_\odot$ corresponding to the exponential  density profile
(dotted curves), for which we have used the fully analytical results of
\cite{Pe88,PeRi}. (Identical results are obtained  by numerical integration.)
The solid lines in Fig.~\ref{f3} refer to a  well-known approximation
(sometimes called ``semianalytical'') to such results, which is obtained  in
the limit $N_e\to 0$ at the Sun surface (it is not exactly so for the
exponential profile, see Fig.~\ref{f1}).  More precisely, the zeroth order
expansion of the hypergeometric functions \cite{Pe88,PeRi} in terms of the
small parameter $z=i\sqrt{2} G_F r_0 N_e(R_\odot)$ $(|z|\simeq 0.16)$ 
gives, for $N_e(0)\to\infty$,  the result $P_\odot\simeq [\exp(\gamma
c^2_\omega)-1]/ [\exp(\gamma)-1]$  with $\gamma=\pi r_0 \delta m^2/ E$
\cite{Pe88},  which leads to the QV prescription discussed in \cite{Fr00}.  The
differences between the solid and dotted curves in Fig.~\ref{f3} are essentially
due to the ``solar border approximation'' ($N_e\to 0$) assumed in the
semianalytical  case; indeed, the differences  would practically vanish if the
exponential density profile, and thus the ``effective'' Sun radius, were
unphysically continued for $r\gg R_\odot$ (not shown). Such limitations of the
semianalytical approximation have been qualitatively suggested by the authors
of \cite{Na00} but, contrary to their claim, our Fig.~\ref{f3} shows explicitly
that the  semianalytical calculation of $P_\odot$ represents a reasonable
approximation to the analytical one for $\omega$ in {\em both\/} octants,
as also verified in \cite{Semi}.

A comparison of the results of Fig.~\ref{f2} (true density) and of
Fig.~\ref{f3} (exponential density) shows that, in the latter case, the
correction term $c^2_\omega-P_\odot$ tends to be somewhat overestimated, in
particular when the semianalytical approximation is used. We have verified that
such bias  is dominantly due to the difference (up to a factor of $\sim 2$)
between  the true density profile and its exponential approximation around
$r/R_\odot\sim 0.8$ (see Fig.~\ref{f1}) and, subdominantly, to the details of
the density profile shape at the border ($r/ R_\odot\to 1$). As a consequence,
the ``exponential profile'' calculation of $P_\odot$ (either semianalytic
\cite{Go00,Fr00} or analytic) tends  to shift systematically the onset of solar
matter effects to lower values of $\delta m^2/E$. For instance, at
$\tan^2\omega\simeq 1$ (maximal mixing), the value $c^2_\omega-P_\odot=0.05$ is
reached at  $\delta m^2/E\simeq 8\times 10^{-10}$ eV$^2$/MeV for the true
density,  and at $\delta m^2/E$ a factor of $\sim 2$ lower for the exponential
density. In order to avoid artificially larger effects at low $\delta m^2/E$ in
neutrino data analyses, one should numerically calculate $P_\odot$ with the
true electron density profile. The difference between the numerical calculation
and the semianalytic approximation is also briefly discussed in \cite{Fr00} for
$\delta m^2/E\leq 10^{-8}$ eV$^2$/MeV (where $P_c\simeq P_\odot$).

Concerning the phase factor $\delta_\odot$, we confirm earlier indications
\cite{Pa90,Pa91} about its smallness, in both cases of true and exponential 
density. In the latter case, the semianalytic approximation gives
\cite{Pe88,Pa90}, for $\delta m^2/E\to 0$, the $\omega$-independent result
\begin{equation}
\delta_\odot\simeq 
L^{-1}\left\{ r_0 
\left[\ln(\sqrt{2}G_F N_e^0 r_0) + \gamma_E \right]-R_\odot\right\}\ ,\\
\end{equation}
where $\gamma_E(\simeq 0.577)$ is the Euler constant. Numerically, 
$\delta^0_\odot\simeq -5.39\times 10^{-4}$, much smaller than $\delta_R$. As
far as $\delta m^2/E\lesssim 10^{-8}$ ($10^{-7}$) eV$^2$/MeV, we find that
$\delta_\odot$ differs from such value by less than $20\%$ (50\%), the exact
difference depending on the value of  $\tan^2\omega$ (not shown). Therefore,
$\delta_\odot$ is an order of magnitude  smaller than $\delta_R$ in the whole
mass-mixing range considered. A similar behavior is found by using the true
density profile $(|\delta_\odot| \lesssim 10^{-3}$ everywhere). As for $P_\odot$,
we find also for $\delta_\odot$ no  significant dependence on the neutrino
production point in the Sun core. 

In conclusion, we find only modest differences between the analytical calculations
of $P_\odot$, based on the exponential density profile, and its semianalytical
approximation. The difference with respect to the numerical calculation of
$P_\odot$,  based on the standard solar model profile, is instead more pronounced,
and leads to a factor of $\sim 2$ difference in the value of $\delta m^2/E$ where QV
effects start to be significant. Therefore, we recommend the use of the electron
profile from the standard solar model, implying numerical calculations of $P_\odot$.%
\footnote{The interested reader can obtain numerical  tables of $P_\odot$,
calculated for the standard solar model density, upon request from the
authors.}
Concerning the phase $\delta_\odot$, it simply shifts the global oscillation phase
$\xi$ in Eq.~(\ref{Pmat}) by less than one permill, and thus it can be safely
neglected in all current applications.

\section{Matter effects in the Earth}
\label{S5}

Strong Earth matter effects typically emerge in the range where  $L_{\rm
osc}\sim L_{\rm mat}$ within the mantle ($N_e \sim 2$ mol/cm$^3$) or the core
($N_e\sim 5$ mol/cm$^3$), as well as in other ranges of mantle-core oscillation
interference \cite{MaCo}, globally corresponding to $\delta m^2/E\sim
10^{-7}$--$10^{-6}$ eV$^2$/MeV. Therefore, only marginal effects are expected
in the parameter range considered in this work, as confirmed by the results
reported in Fig.~\ref{f4}.

Figure~\ref{f4} shows isolines of the quantity $c^2_\omega-P_\oplus$, which
becomes zero in the just-so oscillation limit of very small $\delta m^2/E$. 
The solid curves corresponds to a nadir angle $\eta=0^\circ$  (diametral
crossing of neutrinos) and the dotted curves to $\eta=45^\circ$ (crossing of
mantle only).  For other values of $\eta$ (not shown),  the quantity
$c^2_\omega-P_\oplus$ has a comparable magnitude.  In the current neutrino
jargon, the Earth effect shown in Fig.~\ref{f4}  is operative in the lowermost
part of the so-called ``LOW''  MSW solution \cite{Fo99}  to the solar neutrino
problem, or, from another point of view, to the uppermost part of the  vacuum
solutions \cite{Ba00}. Concerning the phase correction $\delta_\oplus$ (not
shown),  it is found to be smaller than $1.5\times 10^{-5}$ in the whole
mass-mixing plane considered, and thus can be safely neglected.

In practical applications, the correction term $c^2_\omega-P_\oplus$ must be
time-averaged. This poses, in principle, a tedious integration problem, since
such correction appears, in Eq.~(\ref{Pmat}), both in the amplitude of the
oscillating term ($\propto\cos\xi$) and in the remaining, non-oscillating term.
While the integration over  time can be  transformed, for the non-oscillating
term, into a more manageable  integration over $\eta$ \cite{Li97}, this cannot
be done for the oscillating term, which depends on time both through the
prefactor $\sqrt{P_\oplus(1-P_\oplus)}$ and through the phase $\xi$ (via
eccentricity effects). However, in region of the mass-mixing plane  where the
Earth effect is non-negligible (namely, where  $c^2_\omega-P_\oplus
\gtrsim$~few~\%), it turns out that $P_\odot\sim 0$, so that the total
amplitude of the oscillating term is small. Moreover, as it will be discussed 
in the next section, energy smearing effects strongly suppress the oscillation
factor $\cos\xi$ in the same region. Therefore, for $\delta m^2/E\gtrsim
10^{-8}$ eV$^2$/MeV, the oscillating term in Eq.~(\ref{Pmat}) is doubly 
suppressed, and one can safely consider only the non-oscillatory terms.

\section{Damping of the interference term}
\label{S6}

In the just-so regime, the interference factor $\cos\xi$ in Eq.~(\ref{Pmat})
can lead to an observable modulation both in the energy  and in the time
distribution of solar neutrino events. This modulation  gradually  disappears
as $\delta m^2/E$ increases, as a consequence of several decoherence effects 
($\langle \cos\xi\rangle \to 0$) \cite{Cohe,Fr99},   which are typically
dominated by energy smearing \cite{BaFr}. The broader the neutrino spectrum,
the lower the values of $\delta m^2/E$ where decoherence becomes important:
therefore, it suffices to consider the narrowest spectra (the so-called Be
neutrino ``lines''  \cite{Line})  for illustration purposes.

Let us consider a $\nu$ ``line'' energy spectrum $s(E)$, with $\int\! dE\,
s(E)=1$ and $\langle E\rangle =\int \!dE\,s(E)\,E$. The energy-averaged
oscillating factor,
\begin{equation}
C=\left\langle \cos \left(\frac{\delta m^2 L}{2E}\right)
\right\rangle_E = \int \! dE\,s(E)\,\cos\left(\frac{\delta m^2
L}{2E}\right)\ , 
\end{equation}
can be written, in the narrow-width approximation $(\Delta E= E-\langle
E\rangle\ll \langle E\rangle)$, in terms of the Fourier transform of the
spectrum,
\begin{equation}
\tilde s(\tau)= \int\!dE\,s(E)\,e^{i\Delta E\tau}\ . 
\end{equation}
More precisely,
\begin{eqnarray}
C &\simeq& \int \!dE\,s(E)\,
\cos\left(\frac{\delta m^2
L}{2\langle E\rangle} \left(1-\frac{\Delta E}{\langle E\rangle}\right)\right)\\
&=& D  \cos \left(\frac{\delta m^2 L}{2\langle E\rangle}
(1-\delta)\right)\ ,
\end{eqnarray}
where $D=|\tilde s(\tau)|$, $ \delta  = 
\tau^{-1}\langle E\rangle^{-1}\arg \tilde s(\tau)$,
and $\tau=\delta m^2 L/2\langle E\rangle^2$.

Using the $s(E)$ profile for the Be lines from \cite{Line}, we find that the
phase correction $\delta$ is smaller than a few permill in the region where the
damping factor $D$ is greater than a few percent. If $\delta$ is
neglected, the average oscillation factor can be simply written as the
oscillation factor at the average energy, times a damping term $D$:
\begin{equation}
\left\langle \cos \left(\frac{\delta m^2 L}{2E}\right)
\right\rangle_E \simeq D \cdot \cos \left(\frac{\delta m^2 L}
{2\langle E\rangle}\right)\ .
\end{equation}
A similar approach to smearing effects for neutrino lines was developed in
\cite{Pa91}.

Figure~\ref{f5} shows the damping factor $D$ for the two Be neutrino lines at
$\langle E \rangle=0.863$ and $0.386$ MeV. The factor is negligible for $\delta
m^2/E \gtrsim 10^{-8}$ eV$^2$/MeV, implying that the oscillation pattern is
completely smeared out in such range.%
\footnote{This range corresponds  approximately  to $L_{\rm osc}/L \lesssim
w/\langle E\rangle$, where $w\sim O(1 {\rm\ keV})$ is the line width.}
The onset of smearing effects is shifted to even lower values of $\delta m^2/E$
for continuous energy spectra. In conclusion, the integration over neutrino
energy makes the average oscillation factor $\langle\cos\xi\rangle$ always
negligible in the range where Earth matter effects are important, even for
the narrow Be neutrino lines. Such accidental simplification should not make 
one forget that, in general, the transition from coherent to incoherent 
oscillations is a complex phenomenon that deserves further studies (see, e.g.,
\cite{MaKu} for a recent {\em ab initio\/} approach).

Finally, we recall that integration over time acts as a further
damping factor \cite{PeRi}.
At first order in the eccentricity  $\epsilon$, one has that $L(t)\simeq
L[1-\epsilon \cos \alpha(t)]$ and $\dot{\alpha}(t)\simeq 1+2\epsilon
\cos\alpha$, where $\alpha$ is the orbital phase, and $t\in[0,2\pi]$ is the
normalized time variable ($t=0$ at perihelion).  Then the yearly average of
$\cos\xi$ [times the square-law factor $L^2/L^2(t)$] can be performed
analytically \cite{Faid}, giving
\begin{equation}
\frac{1}{2\pi}\int_0^{2\pi} \! dt \;\frac{L^2}{L^2(t)}
\cos\left(\frac{\delta m^2 L(t)}{2E}\right) \simeq  J_0\left(\epsilon\,
\frac{\delta m^2 L}{2E}\right) \cos\left(\frac{\delta m^2 L}{2E}\right)\ ,
\label{faid}
\end{equation}
where $J_0$ is the Bessel function, acting as a further damping term for large
values of its argument.

Notice that the maximum fractional variation of the orbital radius,
$(L_{\max}-L_{\min})/L=2\epsilon =3.34\times 10^{-2}$,  is an order  of
magnitude larger than $\delta_R=R_\odot/L=4.7\times 10^{-3}$ which, in turn,
is larger than the phase corrections $\delta_\odot$ and $\delta_\oplus$.
Therefore, one can safely neglect $\delta_R$, $\delta_\odot$ and
$\delta_\oplus$ in  practical applications involving yearly (or even seasonal)
averages, as we do in this work. However, for averages over shorter time 
intervals, such approximation might break down. In particular,
$\delta_R$ ($\delta_\odot$) might
be comparable to the monthly
(weekly) variations of the solar neutrino signal. The observability of
such short-time variations is beyond the present sensitivity of real-time
solar $\nu$ experiments and  would require, among  other things, 
very high statistics
and an extremely stable level of both the signal detection efficiency and of
the background. If such difficult experimental goals will be reached in  the
future, some of the approximations  discussed so far (and recollected in the
next section) should be revisited and possibly improved.

\section{Practical Recipes}
\label{S7}

We have seen in the previous sections that, as  $\delta m^2/E$ increases, the
deviations of $P_\odot$ (and subsequently of $P_\oplus$) from the vacuum value
$c^2_\omega$ become increasingly important. We have also seen that the phase
corrections $\delta_\odot$ and $\delta_\oplus$ are smaller than
$\delta_R=R_\odot/L$, which can in turn be neglected in present applications,
so that one can practically take the usual vacuum value for the oscillation
phase, $\xi\simeq \delta m^2 L/2E$. We think it useful to organize known and
less known results through the following approximate expressions for the
calculation of $P_{ee}$, which are accurate to better than 3\% with respect to
the exact, general formula (\ref{Pmat}) valid at any $\delta m^2/E$.

For $\delta m^2/E \lesssim 5\times 10^{-10}$ eV$^2$/MeV, one can take
$P_\odot\simeq P_\oplus\simeq c^2_\omega$, and obtain the just-so oscillation
formula
\begin{equation}
P^{\rm JS}_{ee} \simeq c^4_\omega + s^4_\omega + 2s^2_\omega c^2_\omega \cos\xi\ ,
\label{P1}
\end{equation}
with $\xi=\delta m^2 L/2E$. For $5\times 10^{-10}\lesssim \delta m^2/E\lesssim
10^{-8}$ eV$^2$/MeV, one can still take $P_\oplus \simeq c^2_\omega$, but since
$P_\odot\neq c^2_\omega$  (quasi-vacuum regime) one has that 
\begin{equation}
P^{\rm QV}_{ee} \simeq c^2_\omega P_\odot + s^2_\omega (1-P_\odot) + 
2s_\omega c_\omega \sqrt{P_\odot (1-P_\odot)}\cos\xi\ ,
\label{P2}
\end{equation}
where $\xi=\delta m^2 L/2E$, and $P_\odot$ has to be calculated numerically
(see also \cite{Pe88,PeRi,Pa90} for earlier versions of the above equation).

Finally, for $\delta m^2/E\gtrsim 10^{-8}$ eV$^2$/MeV, also Earth matter
effects are important $(P_\oplus\neq c^2_\omega)$; however, this complication
is balanced by the disappearance of the oscillating term  ($\cos\xi \simeq 0$)
due to unavoidable smearing effects, so that  the usual MSW regime is
recovered:
\begin{equation}
P^{\rm MSW}_{ee} \simeq P_\oplus P_\odot +(1-P_\oplus)(1-P_\odot)\ .
\label{P3}
\end{equation}

Time averages are then relatively simple to implement. In  Eqs.~(\ref{P1}) and
(\ref{P2}), the yearly averages of $\cos\xi$ can be performed analytically 
[Eq.~(\ref{faid})]. In Eq.~(\ref{P3}),  the nighttime average of $P_{\oplus}$
can be transformed into a more manageable integration over the nadir angle,
both for yearly \cite{Li97} and for seasonal \cite{Li00} averages.

To summarize,  the above sequence of equations describes the passage from the
regime of just-so to that of MSW oscillations,  via quasi-vacuum oscillations.
In the JS regime, oscillations are basically coherent and do not depend on the
electron density in the Earth or in the Sun ($N_e \to\infty$). In the MSW
regime, oscillations are basically incoherent ($L\to\infty$) and, in general,
depend on the detailed  electron density profile of both the Sun and the Earth.
In particular, in the MSW regime one has to take into account the interplay
between the density profile and the neutrino source distribution profile. The
intermediate QV regime is instead characterized by partially coherent
oscillations (with increasing decoherence as $\delta m^2/E$ increases), and by
a  sensitivity to the electron density of the Sun (but not of the Earth). Such
sensitivity is not as strong as in the MSW regime and, in particular,  QV
effects are independent from the specific $\nu$ production point,  which can be
effectively taken at the Sun center.

For the sake of completeness, we mention that, for high values of  $\delta
m^2/E$ ($\gg 10^{-4}$ eV$^2$/MeV), corresponding to $L_{\rm osc}\ll L_{\rm mat}$ 
in the Sun, the sensitivity to matter effects is eventually lost both in the
Sun and in the Earth  ($P_\odot\simeq P_\oplus\simeq c^2_\omega$), and one
reaches a fourth regime sometimes called of energy-averaged (EA) oscillations,
which is totally incoherent and $N_e$-independent:
\begin{equation}
P^{\rm EA}_{ee} \simeq c^4_\omega + s^4_\omega\ .
\end{equation}
Such regime, which predicts an energy-independent suppression of the solar
neutrino flux, seems to be disfavored (but perhaps not yet ruled out) by
current experimental data on total neutrino rates. In conclusion, for $\delta
m^2/E$ going from extremely low values to infinity, one can identify four
rather different oscillation regimes,
\begin{equation}
	{\rm JS} \to\ {\rm QV} \to {\rm MSW} \to {\rm EA}\ ,
\end{equation}
each being characterized by specific properties and applicable approximations.
Experiments still have to tell us unambiguously which of them truly applies to
solar neutrinos.

\section{Three-flavor oscillation analysis}
\label{S8}

As discussed in \cite{Fr00},   in the QV regime the $2\nu$ survival probability
(\ref{P2}) is non-symmetric with respect to the operation
$\omega\to\frac{\pi}{2}-\omega$, which instead holds for JS oscillations
[Eq.~(\ref{P1})].%
\footnote{In the MSW regime, the mirror asymmetry of the first two
octants was explicitly shown in \protect\cite{Fo96}.}  
Here we extend such observation
to $3\nu$ oscillations, under the hypothesis $\delta m^2\ll m^2$ which leads to
Eq.~(\ref{P3nu}). The  $2\nu$ case is recovered for $\phi=0$.

Equation~(\ref{P3nu}) for $P_{3\nu}$ preserves the (a)symmetry properties  of
$P_{2\nu}$ under the replacement $\omega\to\frac{\pi}{2}-\omega$. Therefore, 
while $P_{3\nu}^{\rm JS}$ [obtained from Eqs.~(\ref{P3nu}) and (\ref{P1})] is
symmetric with respect to the $\omega=\frac{\pi}{4}$ value, the expression of
$P_{3\nu}^{\rm QV}$ [obtained from Eqs.~(\ref{P3nu}) and (\ref{P2})] is not.
Such properties become evident in the triangular representation of the solar
$3\nu$ mixing parameter space discussed in \cite{Fo96,Faid},  to which the
reader is referred for further details. 

Figure~\ref{f6} shows, in the
triangular plot, isolines of  $P_{3\nu}^{\rm QV}$ (dotted lines) for $\delta
m^2/E$ close to  $1.65\times 10^{-9}$ eV$^2$/MeV, corresponding to about 100
oscillation cycles. More precisely, the six panels correspond to 
$\xi=100\times 2\pi+\Delta\xi$, with $\Delta\xi$ from 0 to $\pi$ in steps of
${\pi}/{5}$. The dotted isolines are asymmetric with  respect to 
$\omega=\frac{\pi}{4}$, as is the QV correction $P_\odot-c^2_\omega$. For
increasing $s^2_\phi$ (upper part of the triangle), the symmetry tends to be
restored, mainly because  the asymmetric term $c^4_\phi P_{2\nu}$   in
Eq.~(\ref{P3nu}) is suppressed by the prefactor  $c^4_\phi$ and, marginally,
because the effective electron density $c^2_\phi N_e$ in $P_{2\nu}$ is also
suppressed.%
\footnote{When the effective electron density $c^2_\phi N_e$ is small
($s^2_\phi\to 1$), the statement that the quasi-vacuum probability $P_{2\nu}$
does not depend on the production point in the core (see Sec.~IV) is not
strictly valid, and one should in principle consider also the $\nu$ source
profile.  However, since $P_{2\nu}$ is correspondingly suppressed by
$c^4_\phi$, the source smearing effect is numerically small,  and one can
effectively discard it (e.g., by taking the production point at $r=0$ at any
$s^2_\phi$).}
Exact mirror symmetry at any $\phi$ is restored only if the QV corrections are
switched off (solid lines); in such case,  $P_{3\nu}$ is a quadratic form in
the coordinates $s^2_\phi$ and $s^2_\omega$, and the isolines  are conical
curves.  In any case, $P_{ee}$ becomes typically too low (too high) in the
central region (in the corners) of the triangle where, as a  consequence, one
should expect only marginal solutions to the solar neutrino deficit.

The asymmetry with respect to $\omega=\frac{\pi}{4}$ also shows up in solar
neutrino data fits \cite{Fr00}. Figure~\ref{f7} reports the results of our
global (rates + spectrum + day/night) three-flavor  analysis in the mass-mixing
range $\delta m^2\in [10^{-11},10^{-8}]$ eV$^2$ and 
$\tan^2\omega\in[10^{-2},10^{2}]$, for several representative values of
$\tan^2\phi$. We only show 99\% C.L.\ contours%
\footnote{At 95\% C.L.\ the solutions would be mostly located at $\delta
m^2\lesssim 10^{-9}$ eV$^2$.}
($N_{\rm DF}=3$) for the sake of clarity. The theoretical \cite{BP98} and
experimental  \cite{Cl98,Ha99,Ab99,Fu96,To99,Su99} inputs, as well as the
$\chi^2$ statistical analysis \cite{Stat}, are the same as in Ref.~\cite{Fo99}
(where  MSW solutions were studied). Here, however, the range of $\delta m^2$
is lower, in order to show the smooth transition from the MSW solutions to the
QV and finally the JS ones as $\delta m^2$ is decreased. In particular, the
solutions shown in Fig.~\ref{f7} represent the continuation, at low $\delta
m^2$,  of the LOW MSW solutions shown in Fig.~10 of \cite{Fo99} (panel by
panel).%
\footnote{A technical remark is in order. The minimum value of $\chi^2$ in the
plane of Fig.~\protect\ref{f7} ($\chi^2_{\min}=31.8$) is reached at $(\delta
m^2/{\rm eV}^2,\tan^2 \omega,\tan^2 \phi)= (4.4\times 10^{-10},2.4,0.1)$. For
$\delta m^2>10^{-8}$ eV$^2$ (MSW regime), the minimum value is
$\chi^2_{\min}=27.0$ \protect\cite{Fo99} for the same input data.  In order to
match the results of Fig.~\protect\ref{f7}  in this work and of Fig.~10 in
\protect\cite{Fo99}, we have adopted  the {\em absolute\/} minimum
($\chi^2_{\min}=27.0$) to draw the $\Delta\chi^2=11.34$ contours in
Fig.~\protect\ref{f7}.}
As anticipated in the comments to Fig.~\ref{f6}, the mirror asymmetry  around 
$\tan^2 \omega=1$ decreases for decreasing $\delta m^2$ (JS regime); a little
asymmetry is still present even at $\delta m^2\sim 10^{-10}$ eV$^2$, where the
gallium rates (sensitive to $E$ as low as $\sim 0.2$ MeV) start to feel QV
effects.  In the region where QV effects are important, the solutions are
typically shifted in the second octant $(\omega\gtrsim \pi/4$), since the
gallium rate is suppressed too much in the first octant (see also \cite{Fr00}).
A similar drift was found for the LOW MSW solution \cite{Fo99}.  At any $\delta
m^2$,  the asymmetry decreases at large values of $\phi$ ($\tan^2\phi \gtrsim
1.5$) which, however, are excluded by the combination of accelerator and
reactor data \cite{3atm}, unless the second mass square difference $m^2$ turns
out to be in the lower part of the sensitivity range of the CHOOZ experiment
\cite{CHOO} ($m^2 \sim 10^{-3}$ eV$^2$). For $\phi=0$, the standard two-flavor
case is recovered, and the results are comparable to those found in
\cite{Ba00}. The Super-Kamiokande spectrum plays only a marginal role in
generating the mirror asymmetry of the solutions in Fig.~\ref{f7}, since the
modulation  of QV effects in the energy domain is much weaker than the one
generated by the oscillation phase $\xi$. We find that, at any given $\delta
m^2\lesssim 10^{-8}$ eV$^2$,  the $\chi^2$ difference at symmetric $\omega$
values  is less than $\sim 1$ for the 18-bin spectrum data fit. Therefore, QV
effects are mainly probed by total neutrino rates at present.

We think it is not particularly useful to discuss more detailed features of the
current QV solutions,  such as combinations of spectral data or rates only,
fits with variations of {\em hep\/} neutrino flux, etc.\ (which were instead
given in \cite{Fo99} for MSW solutions). In fact, while the shapes of current
MSW solutions are rather well-defined, those of JS or QV solutions are still 
very sensitive to small changes in the theoretical or experimental input.
Therefore, a detailed analysis of the  ``fine structure'' of the QV solutions
in Fig.~\ref{f7} seems unwarranted at present.

Finally, in Fig.~\ref{f8} we show sections of the allowed $3\nu$ solutions (at 99\%
C.L.) in the triangle representation,  for six selected (increasing) values of
$\delta m^2$. Solutions are absent or shrunk at $s^2_\phi\sim 0.5$, where the
theoretical $\nu$ flux underestimates the gallium and water-Cherenkov data. The
lowest value of $\delta m^2$ ($0.66\times 10^{-10}$ eV$^2$) falls in the JS regime,
so that the ring-like  allowed region  (which resembles the curves of iso-$P_{ee}$
in Fig.~\ref{f6}) is symmetric with respect to the vertical axis at $\omega=\pi/4$.
 However, as $\delta m^2$ increases and QV effects become operative,  the solutions
become more and more asymmetric, and shifted towards the second octant of $\omega$
\cite{Go00,Fr00}.

Figures~\ref{f7} and \ref{f8} in this work, as well as Fig.~10 in \cite{Fo99},
show that solar neutrino data, by themselves, put only a weak upper bound on
the mixing angle $\phi$. Much tighter constraints are set by  reactor data
\cite{CHOO}, unless the second mass square difference  $m^2$ happens to be
$\lesssim 10^{-3}$ eV$^2$ (which seems an unlikely possibility). In any case,
QV effects are operative also for a small (or zero) value of $s^2_\phi$.

\section{Conclusions}
\label{S9}

We have presented a thorough analysis of solar neutrino oscillations in the
``quasi-vacuum'' oscillation regime, intermediate between the familiar just-so
and MSW regimes. The QV regime is increasingly affected by matter effects for
increasing values of $\delta m^2$. We have  calculated such effects both in the
Sun and in the Earth, and  discussed the accuracy of various possible
approximations. We have implemented the QV oscillation probability in a full
three-flavor analysis of solar neutrino data, obtaining solutions which
smoothly join  (at $\delta m^2\sim 10^{-8}$ eV$^2$) the LOW MSW regions found
in \cite{Fo99} for the same input data.  The asymmetry of QV effects makes such
solutions different for $\omega<\frac{\pi}{4}$ and $\omega>\frac{\pi}{4}$, the
two cases being symmetrized only in the just-so oscillation limit of small
$\delta m^2$.

\acknowledgments
We thank J.N.\ Bahcall for providing us with updated standard solar model
results. We thank A.\ Friedland and S.T.\ Petcov for useful discussions.


\begin{figure}
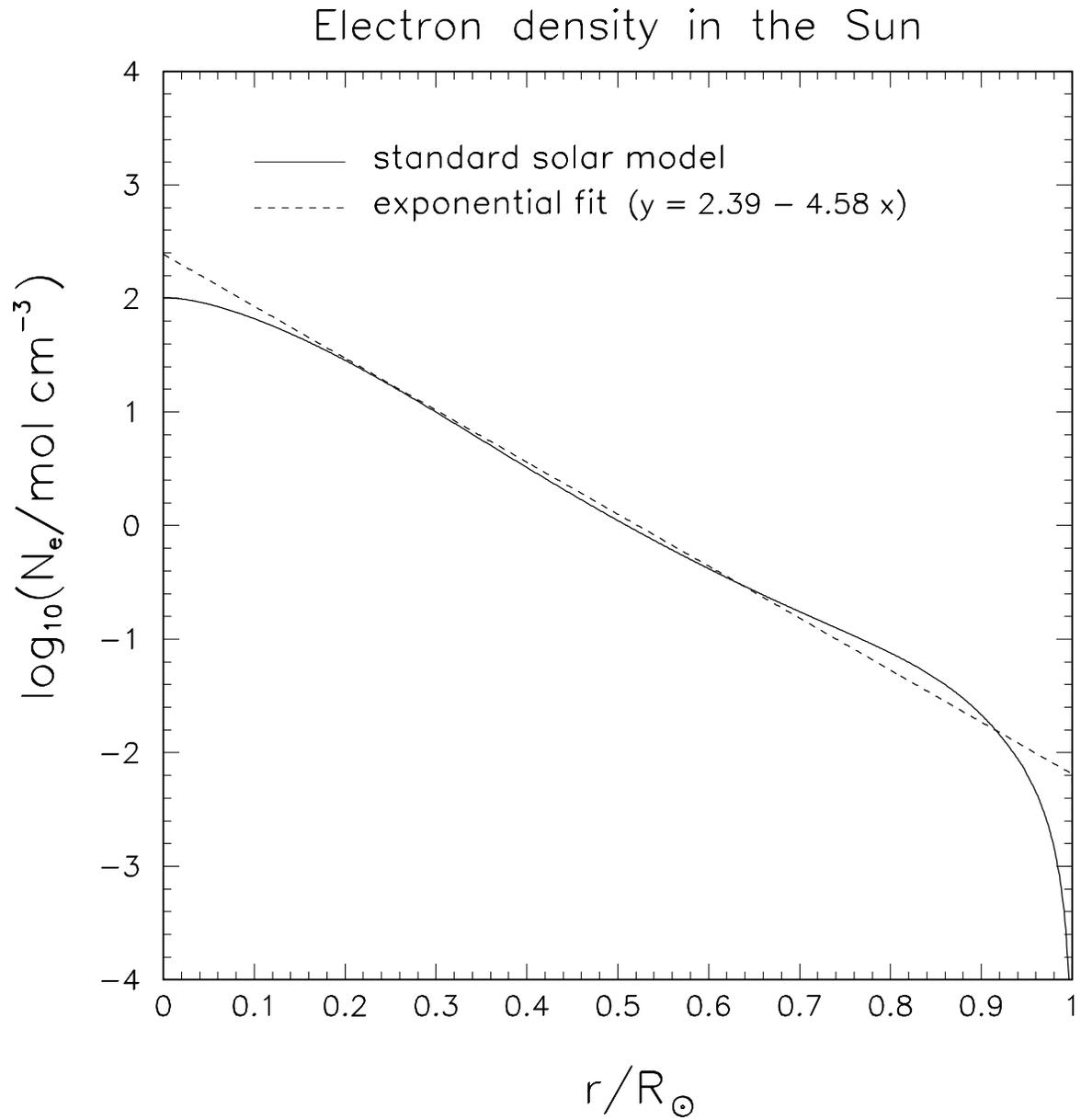

\caption{
Radial profile of the electron density in the Sun from the standard solar model
(solid line), together with its exponential approximation (dashed line). 
\label{f1}}
\end{figure}
\begin{figure}
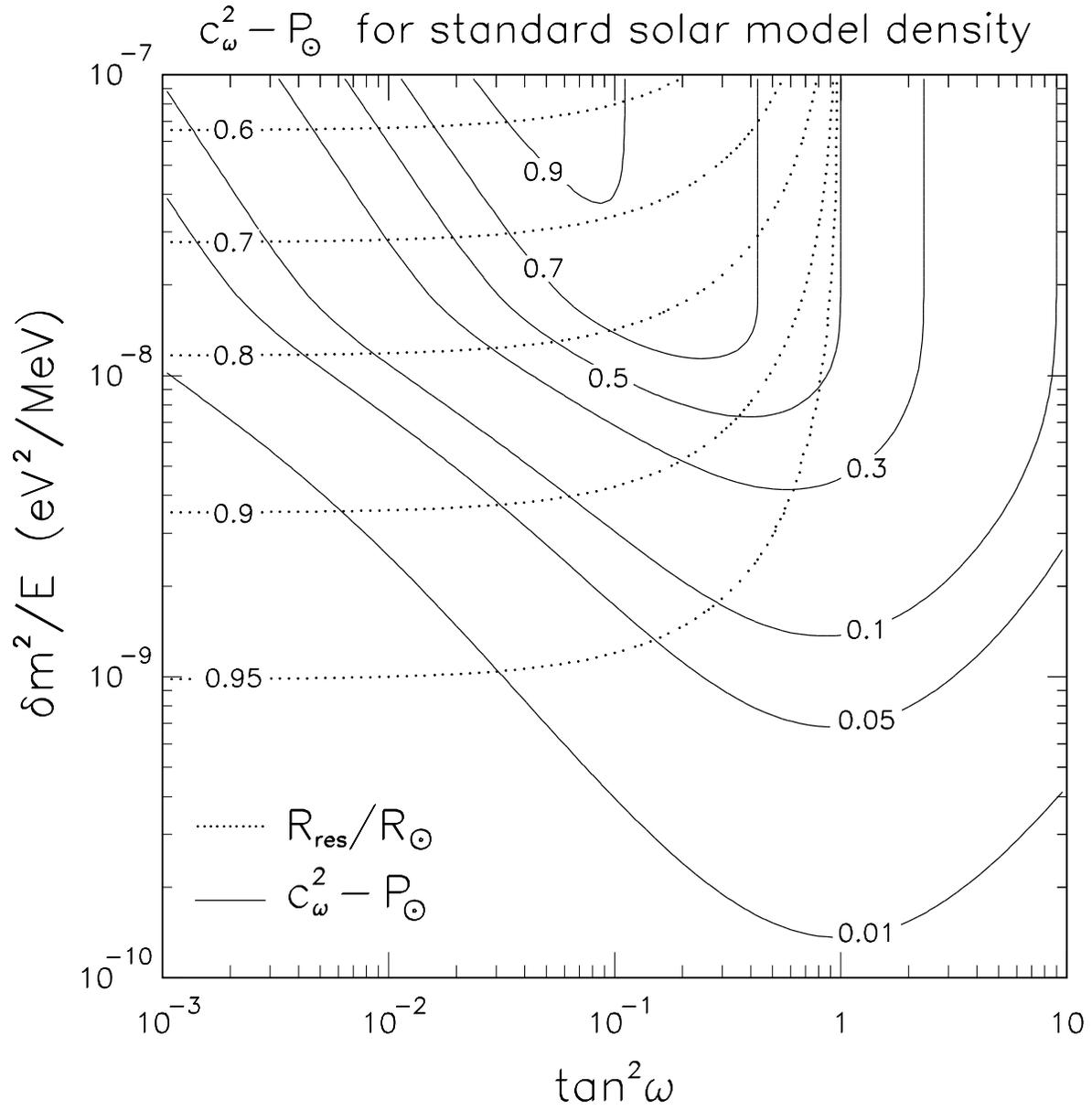

\caption{
Isolines of the solar correction term $c^2_\omega-P_\odot$ in the mass-mixing
plane (solid curves), for standard solar model density \protect\cite{BaPi}. 
Isolines of MSW resonance radii (dashed curves) are also shown. 
\label{f2}}
\end{figure}
\begin{figure}
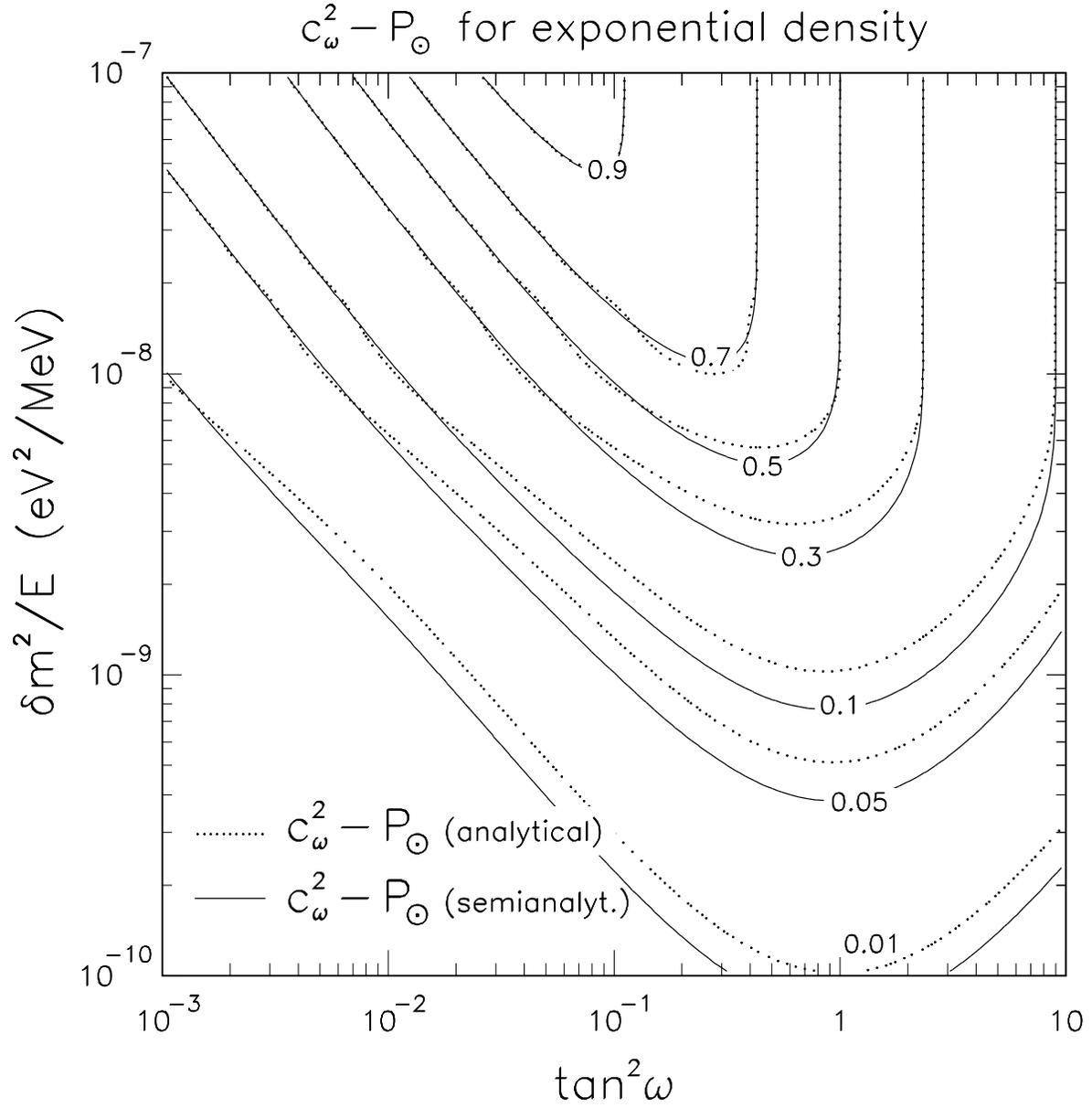

\caption{
As in Fig.~\protect\ref{f2}, but for the exponential density profile. Dotted
and solid lines correspond to analytical calculations and to their
semianalytical approximation, respectively.
\label{f3}}
\end{figure}
\begin{figure}
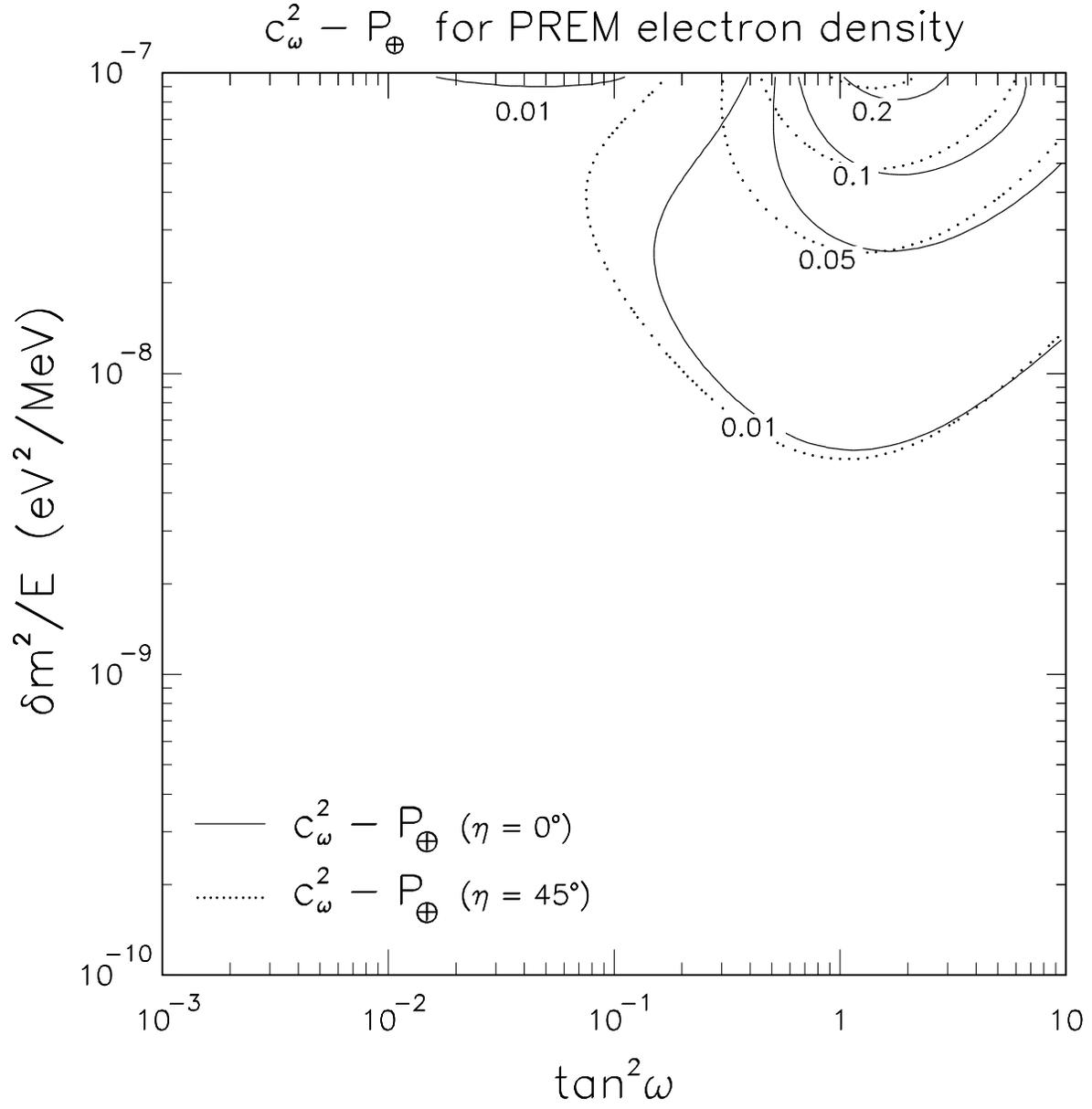

\caption{
Isolines of the Earth correction term $c^2_\omega-P_\oplus$ in the mass-mixing
plane, for PREM \protect\cite{PREM} density. The solid and dotted lines
correspond to a nadir angle equal to $0^\circ$ and $45^\circ$, respectively.
\label{f4}}
\end{figure}
\begin{figure}
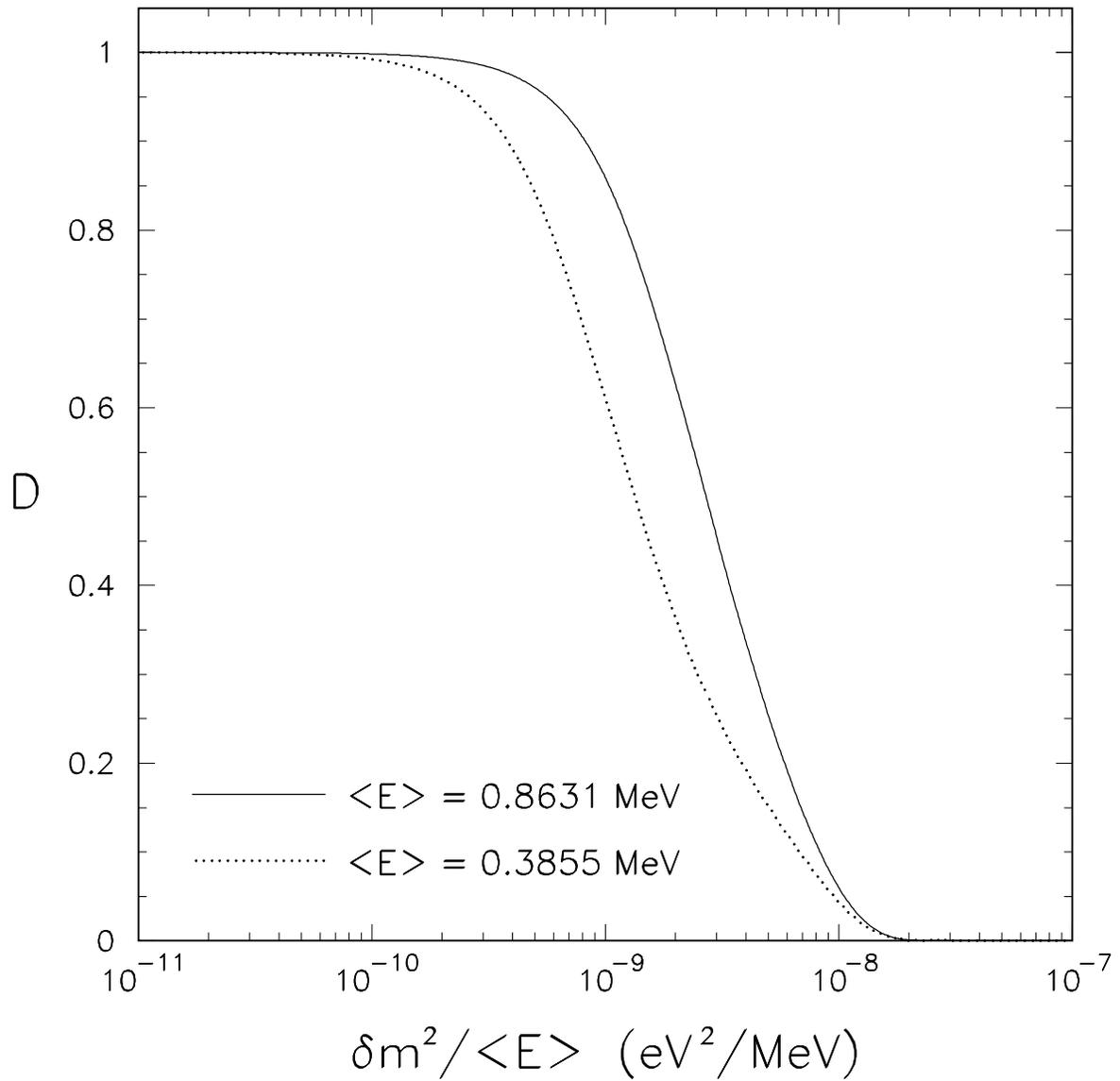

\caption{
Damping factor for the oscillating term, due to the finite Be line width. See
the text for details. 	
\label{f5}}
\end{figure}
\begin{figure}
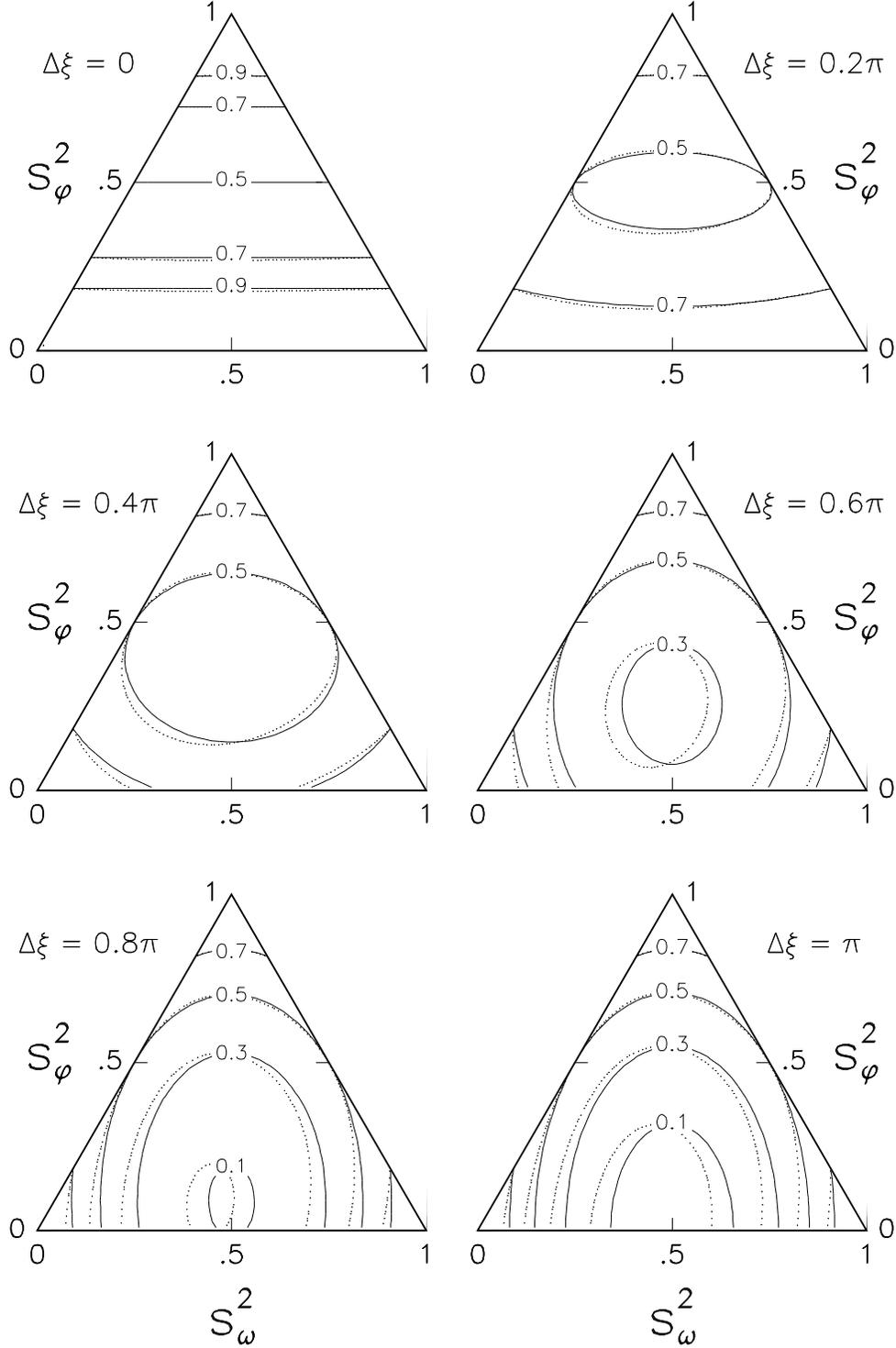

\caption{
Three-flavor oscillations:  Isolines of $P_{ee}$ in the triangular
representation of the solar $\nu$ mixing parameter space, for representative
values of the oscillation phase $\xi$. The dotted and solid lines refer,
respectively, to calculations with and without quasi-vacuum effects. When such
effects are included, the  mirror symmetry $\omega\to\frac{\pi}{2}-\omega$ is
broken.
\label{f6}}
\end{figure}
\begin{figure}
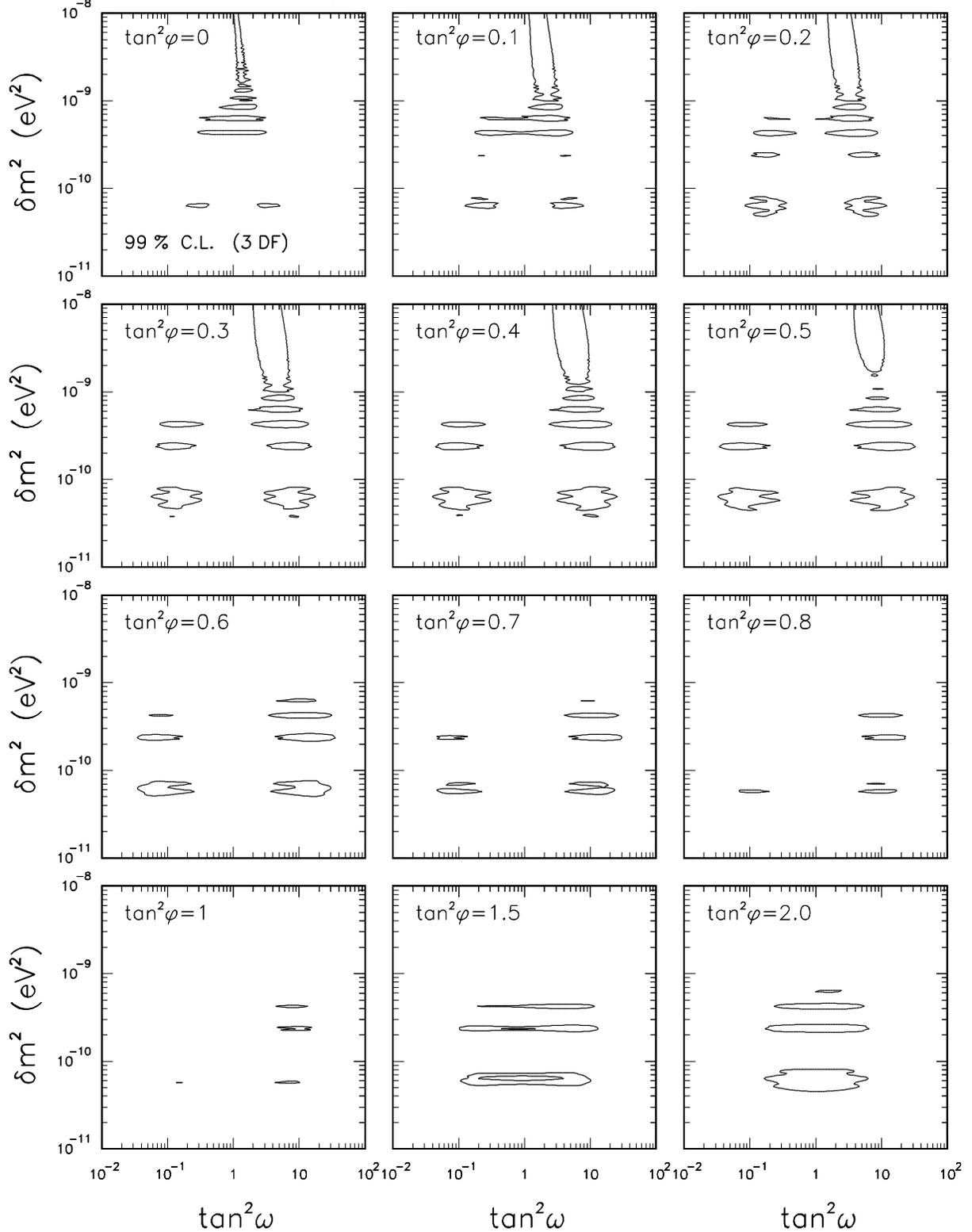

\caption{
Three-flavor oscillations: Contours at 99\% C.L., as derived from a  global
analysis of neutrino data for $\delta m^2\leq 10^{-8}$ eV$^2$ (quasi-vacuum
regime). The asymmetry of the solutions  is evident for increasing values of
$\delta m^2$. The solutions represent the continuation (at low $\delta m^2$) of
the MSW 99\% C.L.\ regions reported in Fig.~10 of \protect\cite{Fo99}.
\label{f7}}
\end{figure}
\begin{figure}
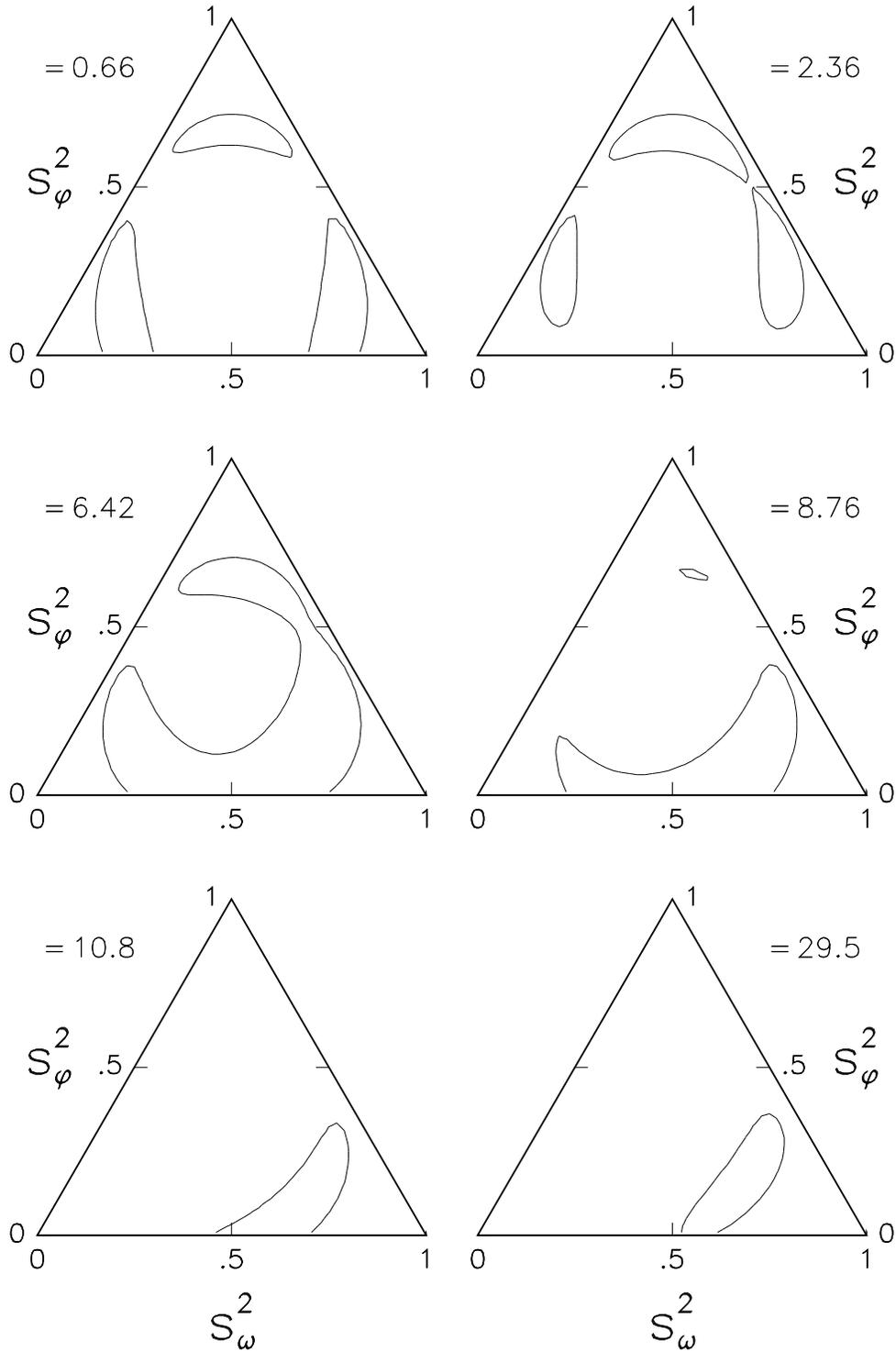

\caption{ 
Sections of the $3\nu$ allowed volume (99\% C.L.) for six  representative
values of $\delta m^2$ (0.66, 2.36, 6.42, 8.76, 10.8, and $29.5\times 10^{-10}$
eV$^2$), shown in the solar triangle plot. 
\label{f8}}
\end{figure}

\newcommand{\InsertFigure}[2]{\newpage\phantom{.}
\vspace*{-2.cm}\begin{center}\mbox{%
\epsfig{bbllx=2truecm,bblly=2truecm,bburx=19.5truecm,bbury=26.5truecm,%
height=23.truecm,figure=#1}}\end{center}\vspace*{-1.85truecm}%
\parbox[t]{\hsize}{\small\baselineskip=0.5truecm\hskip0.5truecm #2}}

\InsertFigure{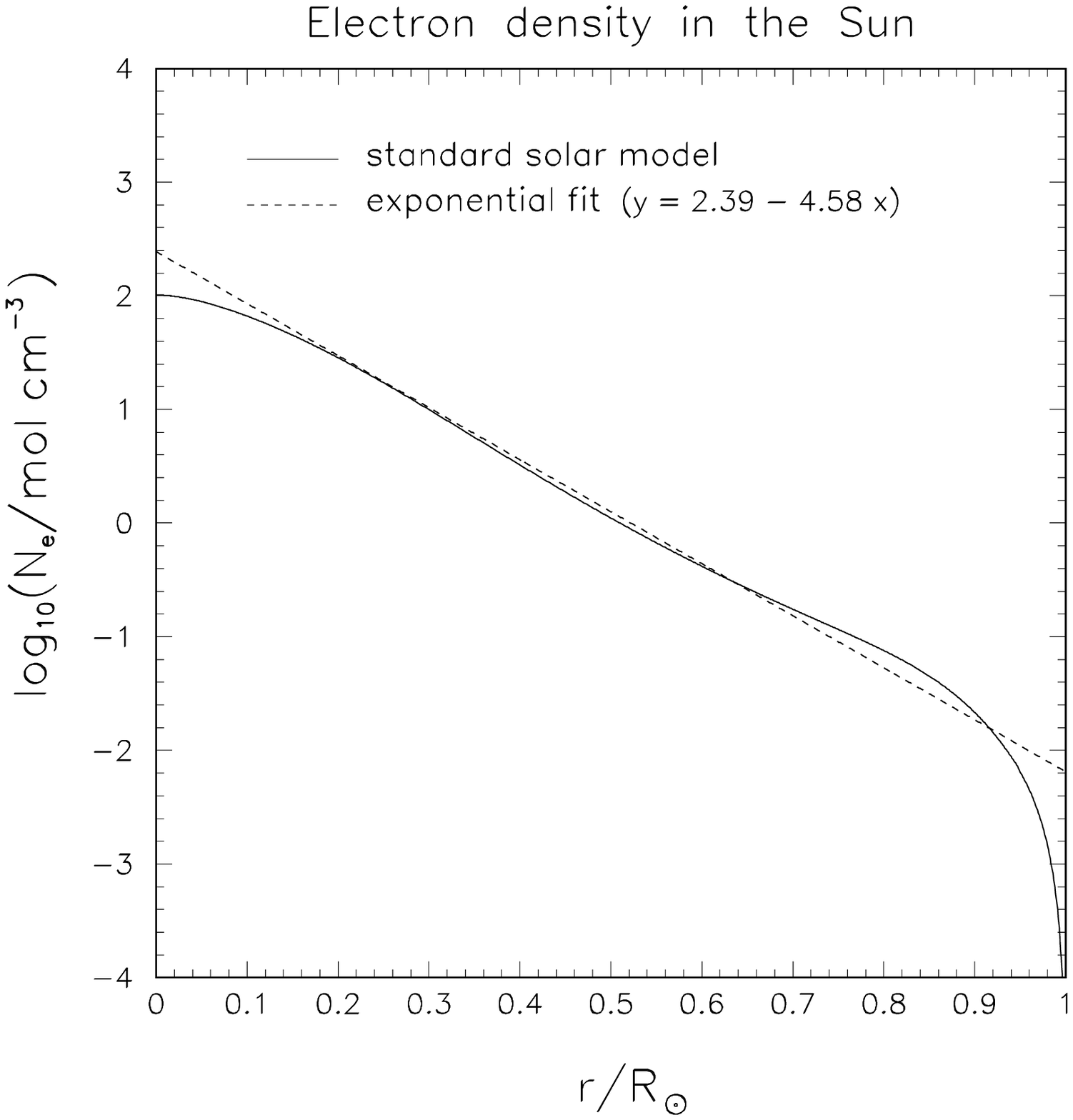}%
{FIG.~\ref{f1}. 
Radial profile of the electron density in the Sun from the standard solar model
(solid line), together with its exponential approximation (dashed line). 
}
\InsertFigure{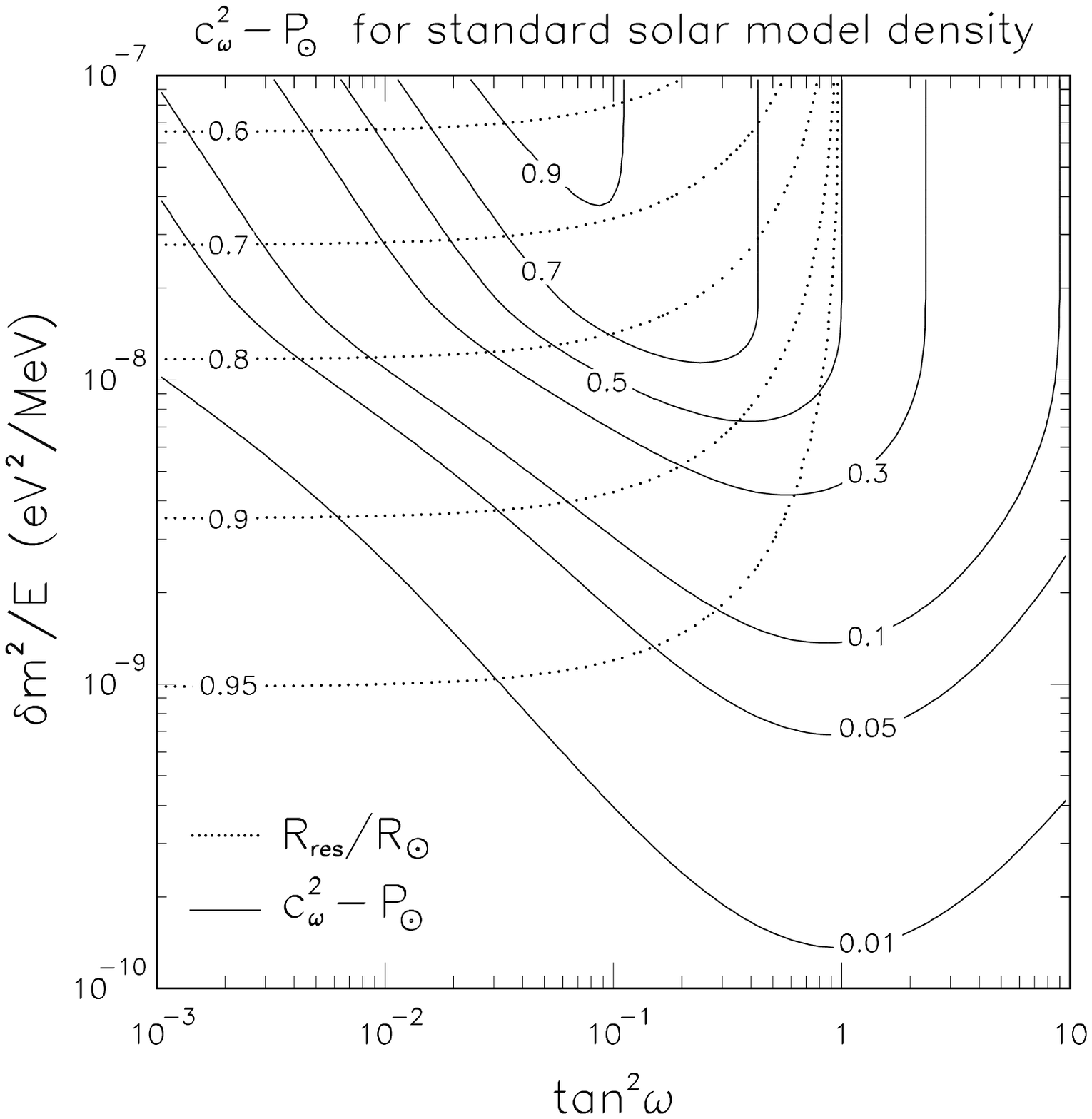}%
{FIG.~\ref{f2}. 
Isolines of the solar correction term $c^2_\omega-P_\odot$ in the mass-mixing
plane (solid curves), for standard solar model density \protect\cite{BaPi}. 
Isolines of MSW resonance radii (dashed curves) are also shown. 
}
\InsertFigure{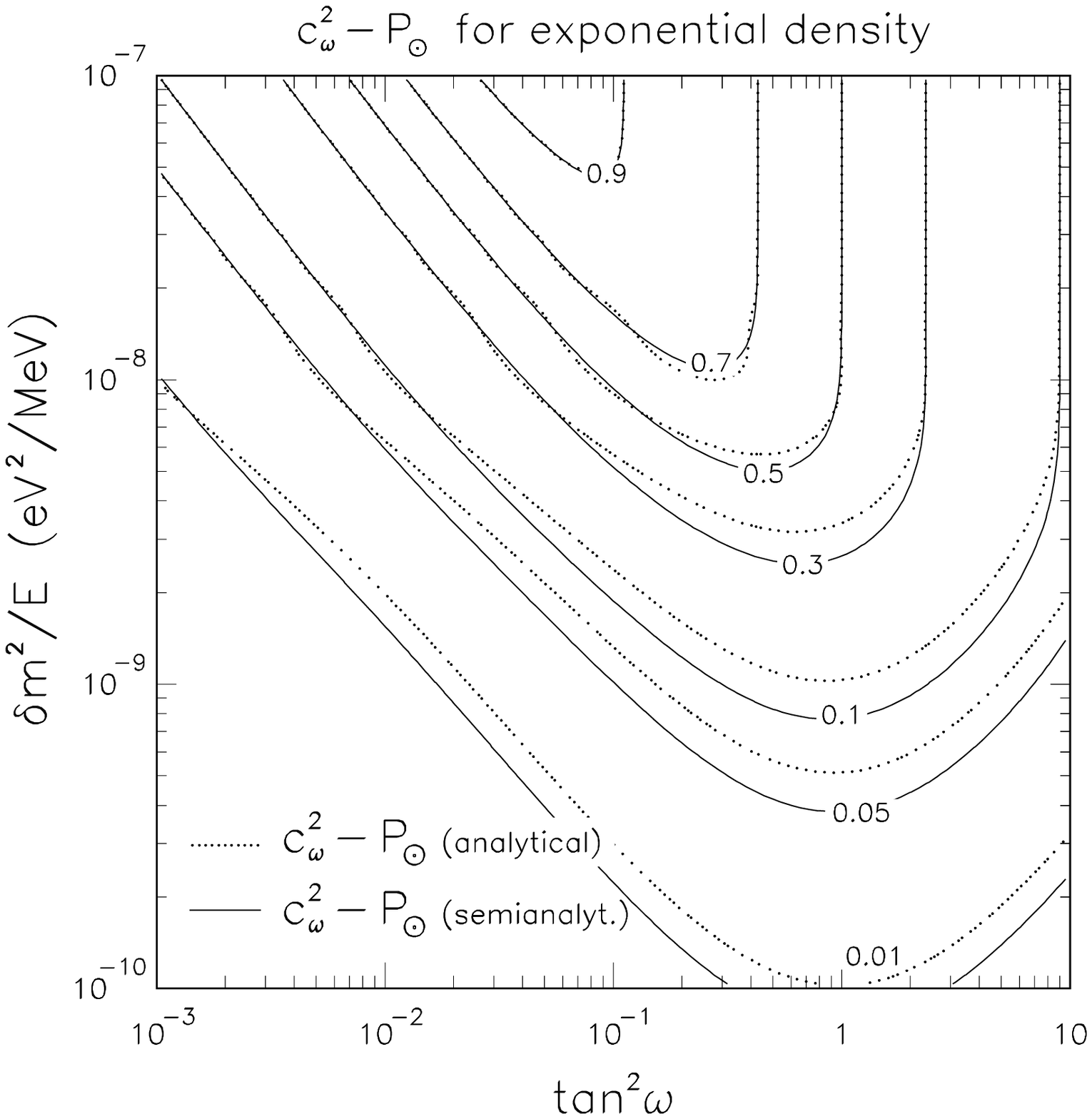}%
{FIG.~\ref{f3}.
As in Fig.~\protect\ref{f2}, but for the exponential density profile. Dotted
and solid lines correspond to analytical calculations and to their
semianalytical approximation, respectively.
}
\InsertFigure{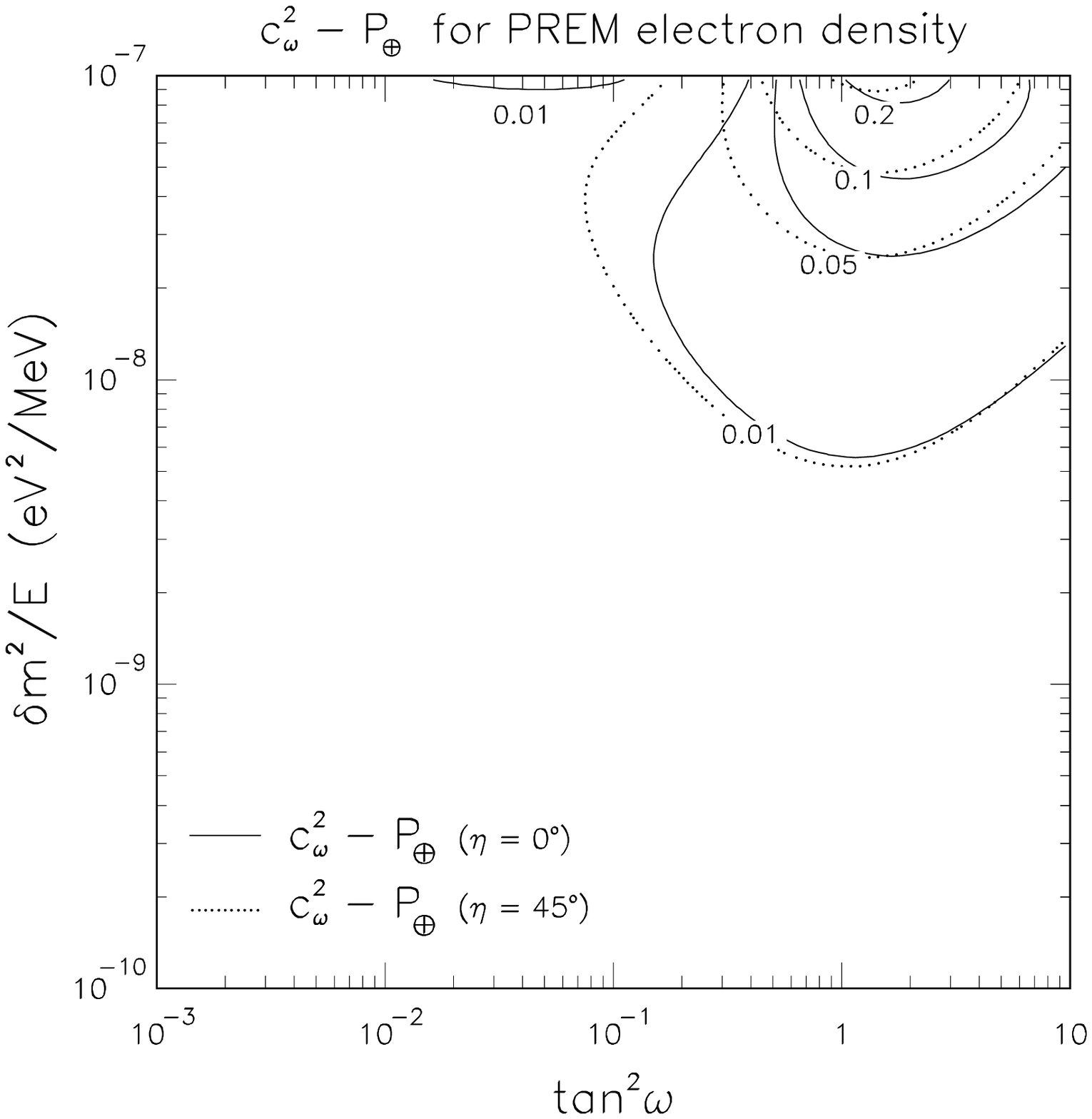}%
{FIG.~\ref{f4}.	
Isolines of the Earth correction term $c^2_\omega-P_\oplus$ in the mass-mixing
plane, for PREM \protect\cite{PREM} density. The solid and dotted lines
correspond to a nadir angle equal to $0^\circ$ and $45^\circ$, respectively.
}
\InsertFigure{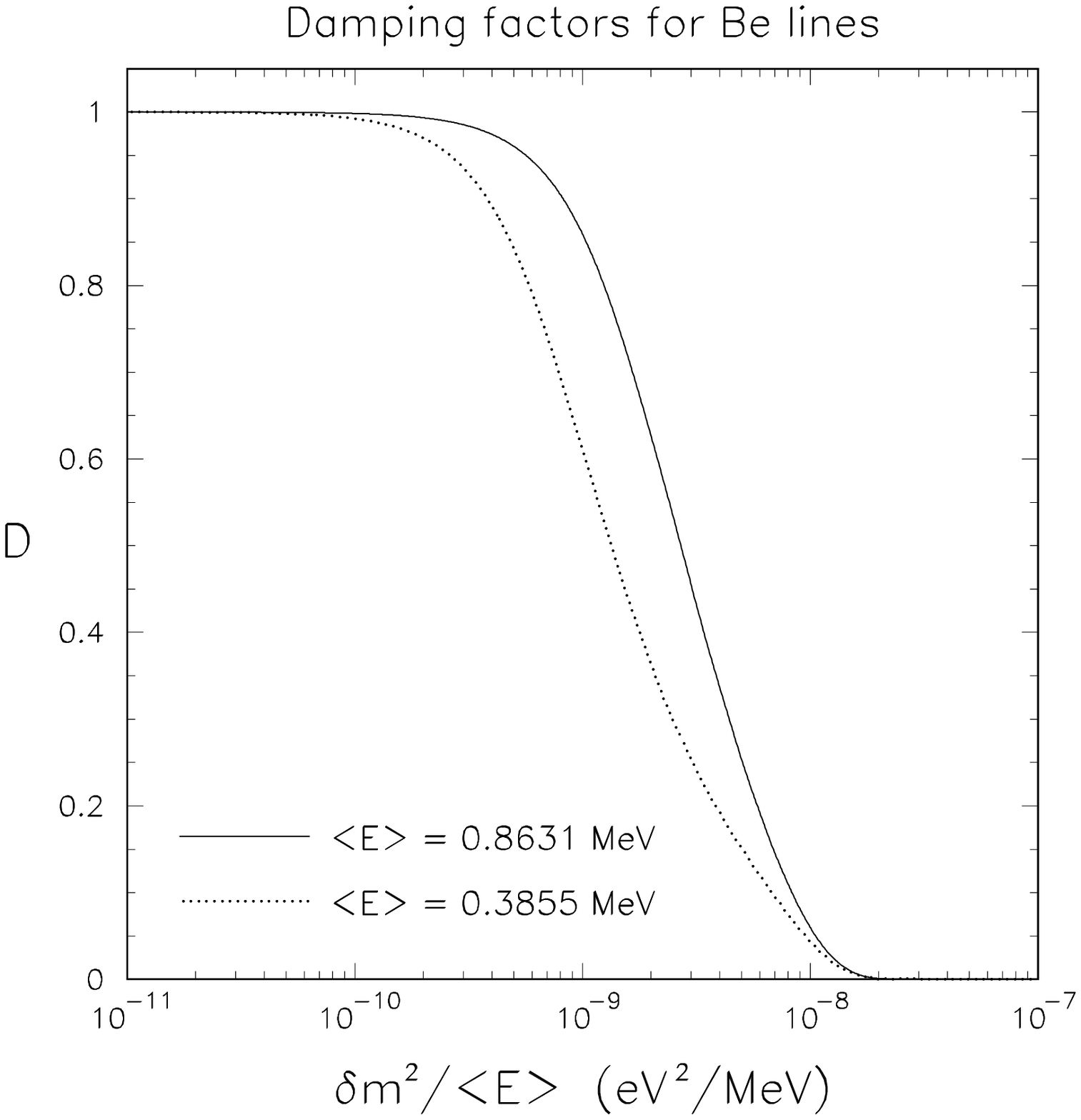}%
{FIG.~\ref{f5}. 
Damping factor for the oscillating term, due to the finite Be line width. See
the text for details. 	
}
\InsertFigure{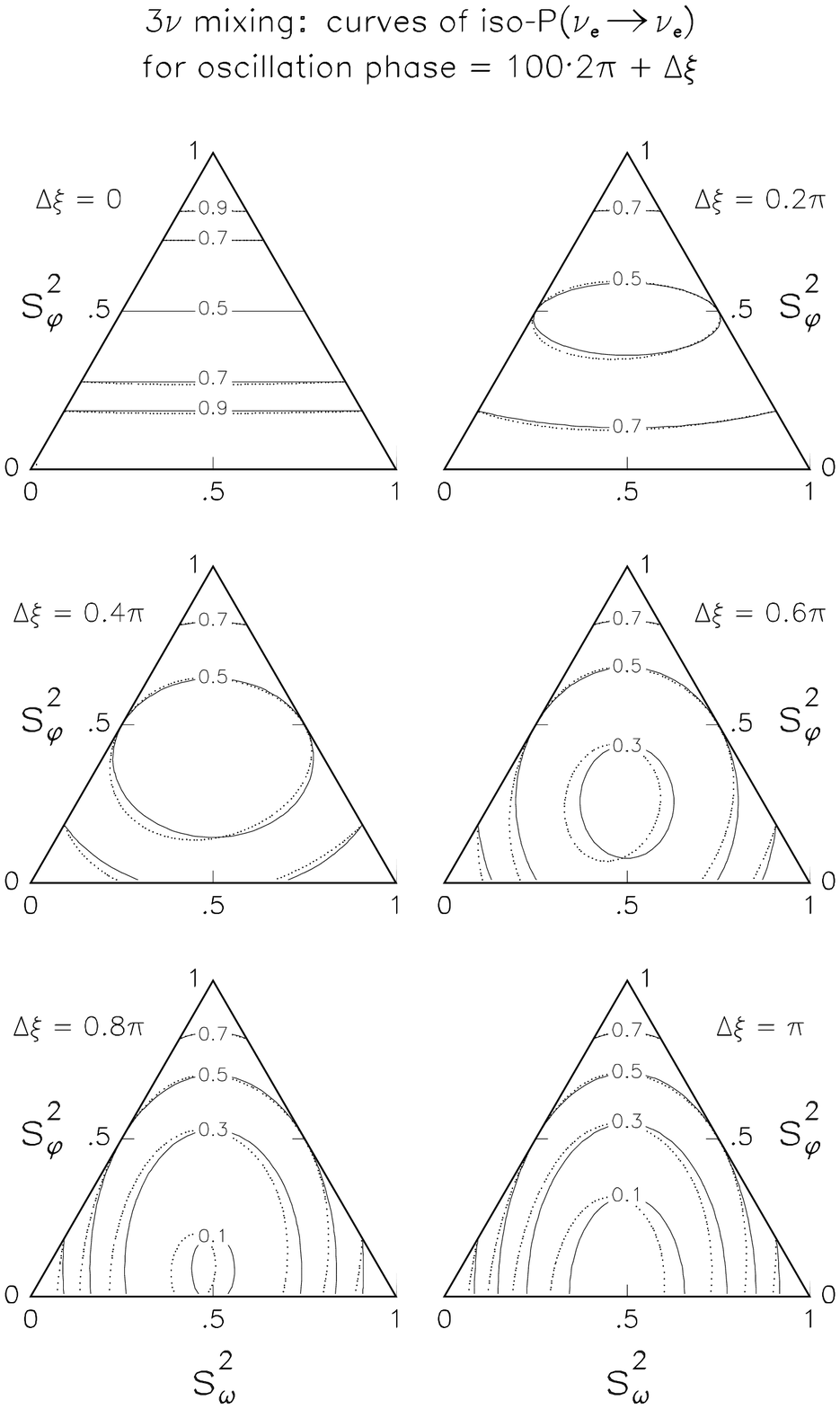}%
{FIG.~\ref{f6}. 
Three-flavor oscillations:  Isolines of $P_{ee}$ in the triangular
representation of the solar $\nu$ mixing parameter space, for representative
values of the oscillation phase $\xi$. The dotted and solid lines refer,
respectively, to calculations with and without quasi-vacuum effects. When such
effects are included, the  mirror symmetry $\omega\to\frac{\pi}{2}-\omega$ is
broken.
}
\InsertFigure{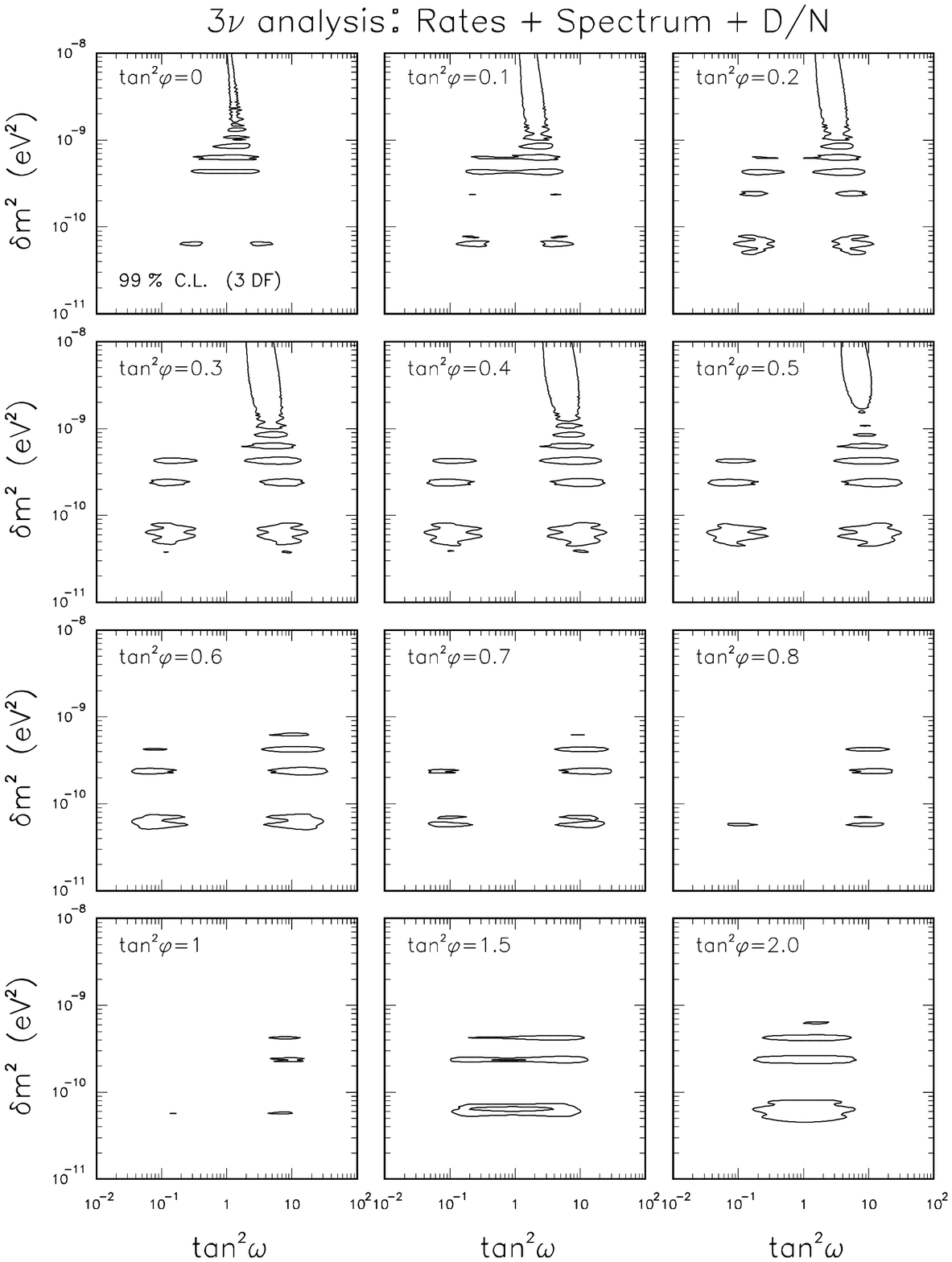}%
{FIG.~\ref{f7}. 
Three-flavor oscillations: Contours at 99\% C.L., as derived from a  global
analysis of neutrino data for $\delta m^2\leq 10^{-8}$ eV$^2$ (quasi-vacuum
regime). The asymmetry of the solutions  is evident for increasing values of
$\delta m^2$. The solutions represent the continuation (at low $\delta m^2$) of
the MSW 99\% C.L.\ regions reported in Fig.~10 of \protect\cite{Fo99}.
}
\InsertFigure{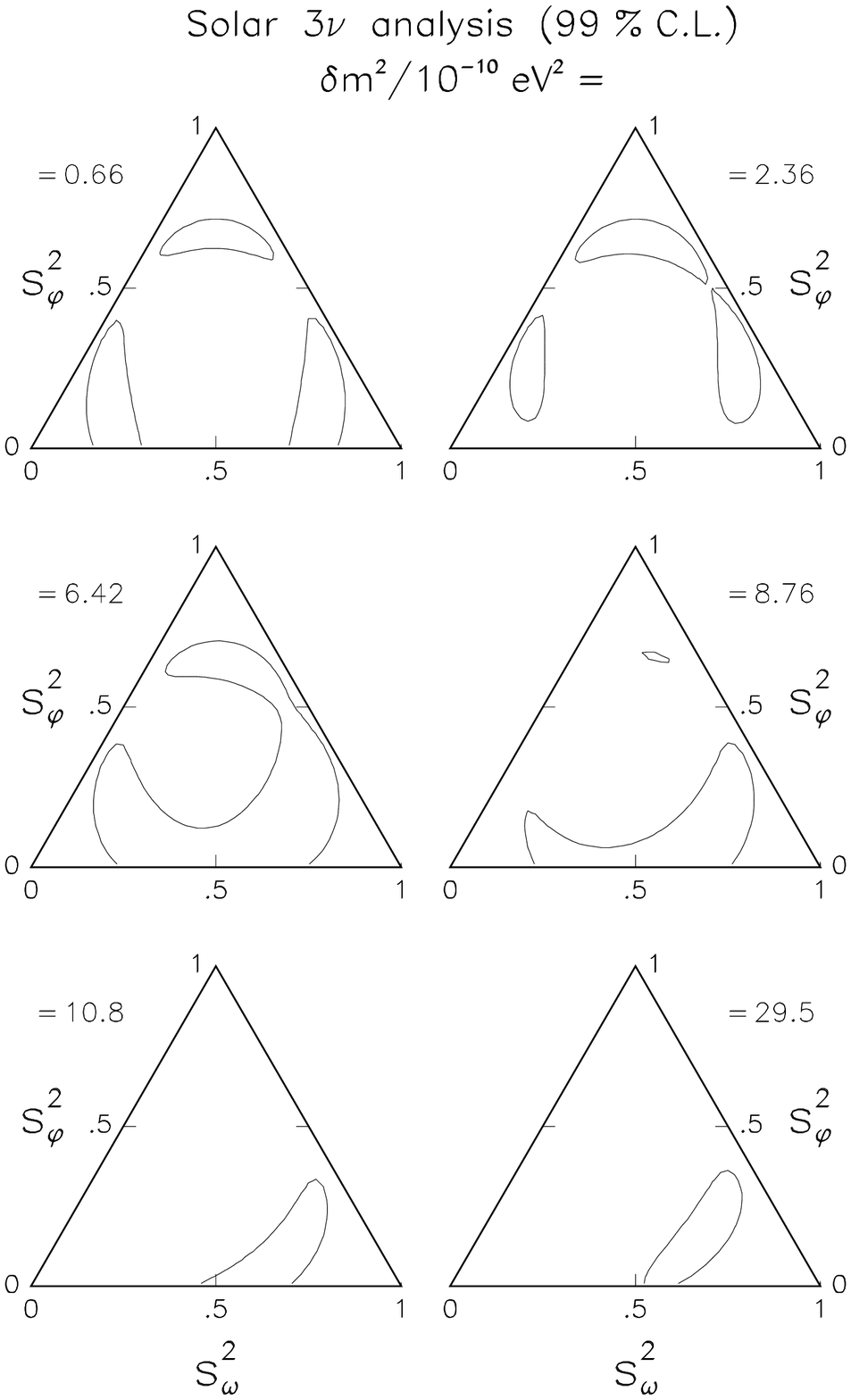}%
{FIG.~\ref{f8}. 
Sections of the $3\nu$ allowed volume (99\% C.L.) for six  representative
values of $\delta m^2$ (0.66, 2.36, 6.42, 8.76, 10.8, and $29.5\times 10^{-10}$
eV$^2$), shown in the solar triangle plot. 
}

\end{document}